\documentclass{SciPost}

\hypersetup{
    colorlinks,
    linkcolor={red!50!black},
    citecolor={blue!50!black},
    urlcolor={blue!80!black}
}

\usepackage[bitstream-charter]{mathdesign}
\usepackage[normalem]{ulem}
\usepackage{orcidlink}
\usepackage{rorlink}

\DeclareSymbolFont{usualmathcal}{OMS}{cmsy}{m}{n}
\DeclareSymbolFontAlphabet{\mathcal}{usualmathcal}

\fancypagestyle{SPstyle}{
\fancyhf{}
\lhead{\colorbox{scipostblue}{\bf \color{white} ~SciPost Physics }}
\rhead{{\bf \color{scipostdeepblue} ~Submission }}

\fancyfoot[C]{\textbf{\thepage}}
}

\begin{document}

\pagestyle{SPstyle}

\begin{center}{\Large \textbf{\color{scipostdeepblue}{
Prospects of future MeV telescopes\\ in probing weak-scale Dark Matter\\
}}}\end{center}

\begin{center}\textbf{
Marco Cirelli\textsuperscript{1$\star$}{\,\orcidlink{0000-0002-4264-6323}} and
Arpan Kar\textsuperscript{1$\dagger$}{\,\orcidlink{0000-0002-2993-3336}}
}\end{center}

\begin{center}
{\bf 1} Laboratoire de Physique Théorique et Hautes Énergies (LPTHE)$^{\ \rorlink{https://ror.org/02mph9k76}}$, CNRS $\&$ Sorbonne Université, 4 Place Jussieu, Paris, France
\\[\baselineskip]
$\star$ \href{mailto:email1}{\small marco.cirelli@gmail.com}\,,\quad
$\dagger$ \href{mailto:email2}{\small arpankarphys@gmail.com}
\end{center}

\section*{\color{scipostdeepblue}{Abstract}}
\textbf{\boldmath{
Galactic weak-scale Dark Matter (DM) particles annihilating into lepton-rich channels not only produce gamma-rays via prompt radiation but also generate abundant energetic electrons and positrons, which subsequently emit through bremsstrahlung or inverse Compton scattering (collectively called `secondary-radiation photons').
While the prompt gamma-rays concentrate at high-energy, the secondary emission falls in the MeV range, which a number of upcoming experiments ({\sc Amego}, {\sc e-Astrogam, {\sc Mast}\dots}) will probe. 
We investigate the sensitivity of these future telescopes for weak-scale DM, 
focusing for definiteness on observations of the galactic center.
We find that they have the potential of probing a wide region of the DM parameter space which is currently unconstrained. Namely, in rather optimistic configurations, future MeV telescopes could probe thermally-produced DM with a mass up to the TeV range, or GeV DM with an annihilation cross section 2 to 3 orders of magnitude smaller than the current bounds, precisely thanks to the significant leverage provided by their sensitivity to secondary emissions. We comment on astrophysical and methodological uncertainties, and compare with the reach of high-energy gamma ray experiments.
}}

\vspace{\baselineskip}

\vspace{10pt}
\noindent\rule{\textwidth}{1pt}
\tableofcontents
\noindent\rule{\textwidth}{1pt}
\vspace{10pt}

\section{Introduction}
\label{sec:introduction}

Despite decades of investigation, Dark Matter (DM) remains one of the most pressing unresolved issues in modern cosmology and particle physics~\cite{Cirelli:2024ssz}. 
Weak-scale particle Dark Matter (DM), which is here broadly defined as having a mass between a few GeV 
and tens of TeV's, remains a motivated framework for the solution of the DM problem. This used to be inspired mostly by the fact that several New Physics theories predicted a weak-scale particle with the appropriate properties to play the role of DM. Nowadays, as the promised New Physics has not showed up (at least yet), the motivation mostly shifted to the fact that particles with weak-scale mass and interactions are produced in the right amount in the early Universe. 
Weak-scale DM is still therefore the subject of intense searches with many different methods~\cite{Cirelli:2024ssz}. 

Indirect Detection (ID) techniques aim at detecting the signals of DM particle annihilations or decays in the Galaxy or in different astrophysical environments. For weak-scale DM, and restricting to the signals in photons, the observables can consist of mainly two things. On one side, high-energy $\gamma$-rays are produced directly in the annihilation (or decay) process, hence called {\em prompt emission}. 
On the other side, lower-energy X-rays/$\gamma$-rays are produced by the electrons and positrons (generated by DM annihilations or decay), in particular via Inverse Compton Scattering (ICS) processes on the ambient light and via bremsstrahlung processes on the galactic gas. These are dubbed {\em secondary emission}. 
In most cases, prompt and secondary emissions both exist, and their relative importance depends on the details of the DM model (for instance, lepton-rich annihilation channels will produce abundant $e^\pm$ and thus a significant ICS signal), on the ambient conditions (for instance, regions of the Galaxy where the ambient light is more intense will be a more favorable target to search for ICS photons) as well as on the sensitivity of the dedicated experiments. 

\medskip

In the past, the prompt emission signal from weak-scale DM received by far most of the attention, although the importance of secondary emission in DM indirect detection searches has been highlighted in a number of studies, see e.g.~\cite{Baltz:2004bb,Colafrancesco:2005ji,Colafrancesco:2006he,Cholis:2008wq,Zhang:2008tb,Cirelli:2009vg,Borriello:2009fa,Barger:2009yt,Cirelli:2009dv,Zavala:2011tt,Beck:2015rna} for ICS and \cite{Cirelli:2013mqa} for bremsstrahlung.
One potential difficulty for detecting secondary emission is the relatively poor sensitivity of past and existing telescopes in the range 0.1-100 MeV, the so-called {\em MeV gap}. Indeed, secondary photons from weak-scale DM will in general fall in the gap.\footnote{For instance, we recall that, as a rule of thumb, ICS processes upscatter the ambient photon energy from its initial low value $E^0_\gamma$ to a final value of up to $E_\gamma \approx 4 \gamma^2 E^0_\gamma$. Here $\gamma = E_e/m_e$ is the relativistic factor of the electrons and positrons. 
Hence a 10 GeV electron will produce a $\sim$ 0.15 MeV hard X-ray when scattering off the CMB ($E^0_\gamma \approx 10^{-4}$ eV), 
or a $\sim$ 1.5 GeV $\gamma$-ray when scattering off optical starlight ($E^0_\gamma \approx 1$ eV). 
For bremsstrahlung, the energy of the emitted photon peaks at a fraction of the initial energy of the $e^\pm$, typically between 1/10 and 1/2, depending on  both the $e^\pm$ spectrum and on local conditions. See sections \ref{sec:secondary_ICS} and \ref{sec:secondary_brem} for the full treatment.}
Fortunately, a number of upcoming or planned MeV telescopes aim at filling this gap \cite{Engel:2022bgx}. 
Among them, {\sc Amego}~\cite{AMEGO:2019gny,2020SPIE11444E..31K,Caputo:2022xpx}, 
{\sc e-Astrogam}~\cite{e-ASTROGAM:2016bph,e-ASTROGAM:2017pxr} 
and {\sc Mast}~\cite{Dzhatdoev:2019kay} are expected to provide a good sensitivity. 
Our main goal in this work is to explore the potential that these future MeV telescopes have in constraining weak-scale DM, compared in particular to high-energy $\gamma$-ray telescopes.

\medskip

Recently, a few studies have done work related to this paper. 
Refs. \cite{Bartels:2017dpb,Cirelli:2020bpc,Cirelli:2023tnx,DelaTorreLuque:2023olp} 
have applied the same principle to another range of DM masses, namely sub-GeV DM. 
In \cite{Boddy:2015efa,Caputo:2022dkz,ODonnell:2024aaw,Saha:2025wgg} the authors 
assess the sensitivity of several future MeV telescopes to MeV-GeV DM, including prompt emission only, 
and \cite{Ray:2021mxu, Cheung:2025gdn} consider the case of Primordial Black Hole (PBH) dark matter. 
In \cite{Eisenberger:2023tgn} the authors consider secondary emission signals for a specific weak-scale DM model and for a specific target (dwarf galaxies). Our work therefore significantly extends and generalizes these studies.

\medskip

The remainder of the paper is organized as follows. 
In section \ref{sec:signals} we discuss the MeV-GeV photon fluxes, focusing in particular on the inner galactic region. 
Section~\ref{sec:DMsignal} provides the discussion 
on the weak-scale DM-induced photon signals in the 
MeV-GeV energy range, focusing on the prompt 
(section~\ref{sec:prompt}) and the secondary 
(sections~\ref{sec:secondary_ICS} and \ref{sec:secondary_brem}) 
emissions. Section~\ref{sec:background} summarizes 
different photon backgrounds 
and data in the MeV-GeV range. 
In section~\ref{sec:future_telescopes} we briefly 
review the different near-future space-based MeV telescopes 
considered in this work. Section~\ref{sec:methodology} details 
the methodology used to derive the 
bounds and the projected sensitivities of the 
MeV telescopes on DM annihilation. 
We then present and discuss the results of our analysis 
in section~\ref{sec:results}; 
{we also discuss the effects of propagation of $e^\pm$ in the Galaxy.} 
Our conclusions are contained in section~\ref{sec:conclusions}. 
Finally, in Appendix~\ref{sec:astrophysical_models} we provide 
a discussion on the impact of varying different 
astrophysical ingredients.

\section{MeV-GeV photons in the Galaxy}
\label{sec:signals}

The full energy range of the observed photons that we consider here is 
$0.1 \, {\rm MeV} \lesssim E_\gamma \lesssim 100 \, {\rm GeV}$, and, in particular, for the 
future MeV telescopes we use the range $0.2 \, {\rm MeV} \lesssim E_\gamma \lesssim 5 \, {\rm GeV}$ 
(for which the galactic background models are provided in \cite{ODonnell:2024aaw}). 
As for the target of observation, we consider a disk of $10^{\circ}$ radius 
around the galactic Center (GC) (following \cite{ODonnell:2024aaw, Coogan:2021sjs, Coogan:2020tuf}), 
which becomes our region of interest (ROI). Such an angular size of observation is of the same order 
as the maximum angular width of the upcoming MeV $\gamma$-ray telescopes 
\cite{AMEGO:2019gny, 2020SPIE11444E..31K}. 
Below we discuss the possible DM signals as well as 
the background photons and observed data from this region 
in our photon energy range of interest.

\subsection{Photon signals from DM annihilation in the Galaxy}
\label{sec:DMsignal}

In this work we assume that the origin of the signal is due to 
the pair-annihilations of 
a single-component self-conjugate weak-scale DM, that explains the entire 
observed DM density. 
We consider the following annihilation channels one at a time, 
assuming a 100\% branching fraction each time: 
${\rm DM \, DM} \rightarrow \mu^+\mu^-$, ${\rm DM \, DM} \rightarrow e^+e^-$, 
${\rm DM \, DM} \rightarrow b\bar{b}$ and ${\rm DM \, DM} \rightarrow W^+W^-$. 
 
We assume that the density distribution of DM in the halo, for the target region, 
is described by the NFW profile~\cite{Navarro:1995iw}: 
\begin{equation}
\rho_{\rm DM}(r) = \frac{\rho_0}{\left(\frac{r}{r_s}\right) \left(1 + \frac{r}{r_s}\right)^2} \, ,
\label{eq:rho_profile}
\end{equation}
where $r$ is radial distance from the GC. 
For the parameters $\rho_0$ and $r_s$ we use the values corresponding to the 
fitted NFW profile parameters (the central values) obtained in \cite{2019JCAP...10..037D} for the  baryonic model B2; they are similar to the ones used in 
\cite{ODonnell:2024aaw, Coogan:2021rez, Coogan:2020tuf}~\footnote{We checked that, 
using the NFW parameters tabulated in \cite{Cirelli:2024ssz}, 
our results (i.e., the DM annihilation signals and the corresponding constraints 
on DM annihilation) change at most by $\sim20 \%$.}. 
In the following, we derive our main results using this NFW profile.  
However, for comparison we will also consider other choices, 
see sec.~\ref{sec:appendix_DM_halo}.  

\medskip

As mentioned in the Introduction, for the considered weak-scale DM scenario, 
the main sources of photon signals from the GC region over the energy range of interest 
($0.1 \, {\rm MeV} \lesssim E_\gamma \lesssim \\ 100 \, {\rm GeV}$) are 
the prompt $\gamma$-ray emission and 
the emissions via Inverse Compton Scatterings and bremsstrahlung. 
As we will see, in some cases, the secondary components, 
especially the ICS, can become more 
important compared to the primary one over the considered photon energy range. 
Below we describe the production of these photon signals~\footnote{Apart from these, 
the production of another secondary photon signal 
from the In-flight annihilation (IfA) of DM induced positrons is also possible \cite{Bartels:2017dpb}. 
However, we checked that, in our considered photon energy range, 
and for our DM scenario, this flux is well suppressed compared to the other types of flux.}.

\subsubsection{Primary signal: prompt $\gamma$-ray emission}
\label{sec:prompt}

DM annihilation into a given channel (e.g., $e^+e^-$, $\mu^+\mu^-$, $b\bar{b}$ or $W^+W^-$) gives rise to the 
prompt $\gamma$-ray flux which at the observer location can be computed as: 
\begin{equation}
\frac{d\Phi_{\rm prompt}}{dE_\gamma d\Omega} = \frac{\langle \sigma v \rangle}{8 \pi \, m^2_{\rm DM}} \, \frac{dN_\gamma}{dE_\gamma} \, 
\frac{J_{\Delta \Omega}}{\Delta \Omega} \, ,
\label{eq:prompt_gamma}
\end{equation}
where $\langle \sigma v \rangle$ and $m_{\rm DM}$ are the velocity-averaged annihilation cross-section and the mass of the DM, respectively. 
The distribution $\frac{dN_\gamma}{dE_\gamma} (E_\gamma, m_{\rm DM})$ denotes the photon energy spectrum produced 
per annihilation in the considered annihilation channel. 
Such a spectrum is determined using the analytical expressions in \cite{Cirelli:2020bpc} for $m_{\rm DM} < 5$ GeV 
and using the {\tt PPPC4DMID} tools~\cite{PPPC_web} for $m_{\rm DM} \ge 5$ GeV. 
Finally, the astrophysical $J$-factor for DM annihilation, $J_{\Delta \Omega}$, 
is defined for the observation region $\Delta \Omega$ as:
\begin{equation}
J_{\Delta \Omega} = \int_{\Delta \Omega} d\Omega \, \int_{l.o.s.} ds \,\, \rho^2_{\rm DM} (r(s,\theta)) \, \, .
\end{equation}
Here $s$ is the line-of-sight ($l.o.s.$) coordinate (with respect to the observer) 
which is related to the radial and angular distances from the GC, $r$ and $\theta$, as: 
$r = \sqrt{s^2 + r^2_\odot - 2 s r_\odot \cos \theta}$, 
with $r_\odot$ being the distance of the Earth from the GC \cite{2019JCAP...10..037D}. 

\begin{figure*}[ht!]
\includegraphics[width=0.5\textwidth]{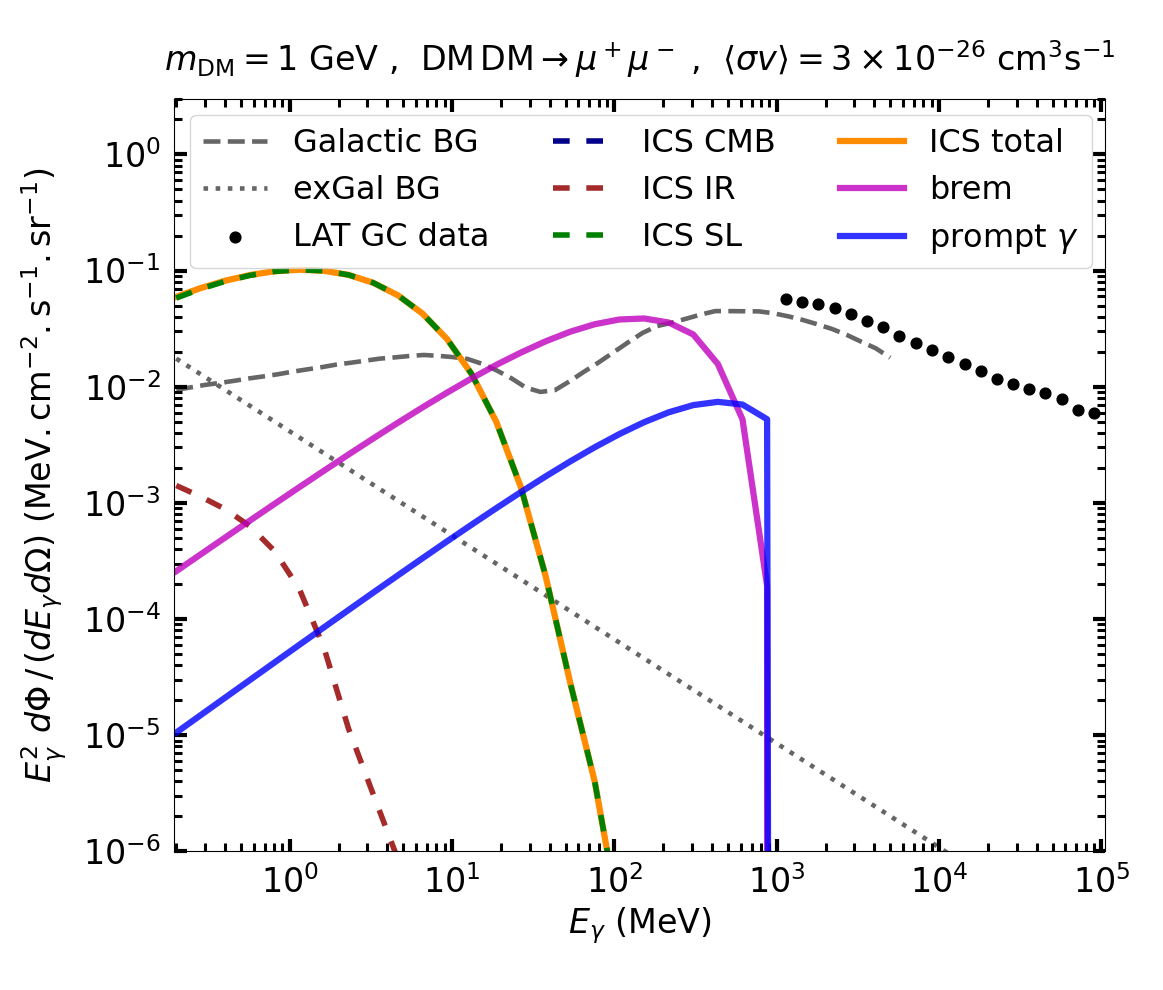}
\includegraphics[width=0.5\textwidth]{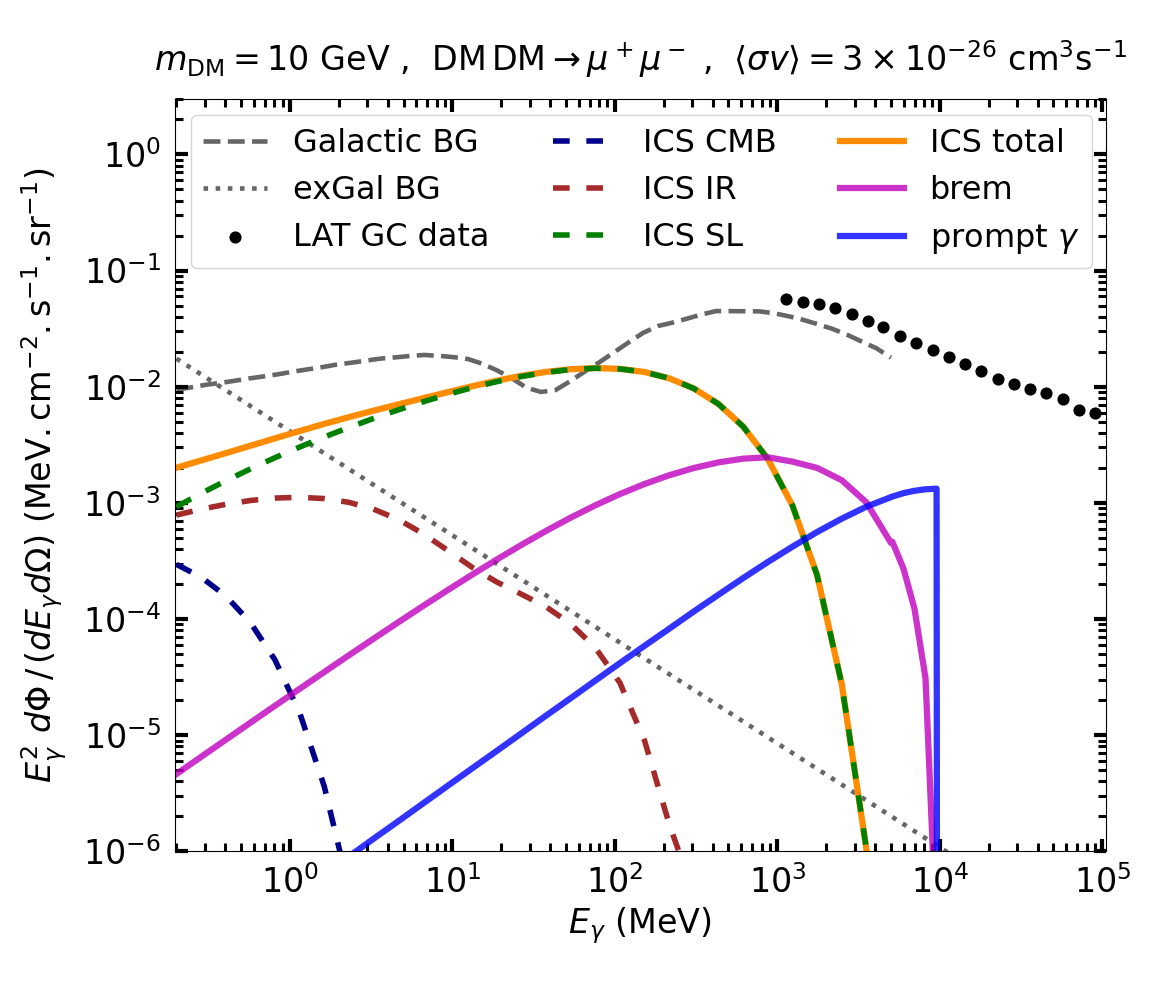}\\
\includegraphics[width=0.5\textwidth]{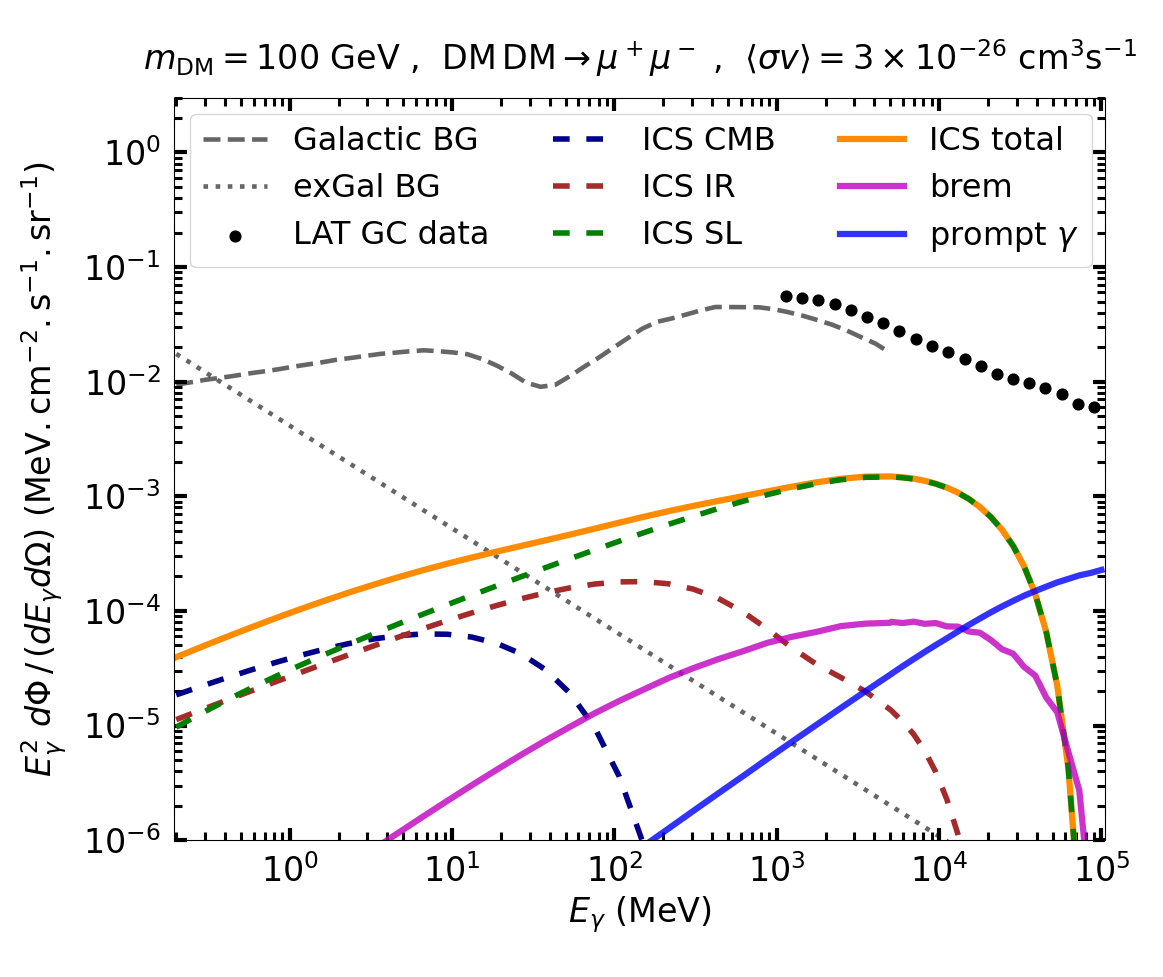}
\includegraphics[width=0.5\textwidth]{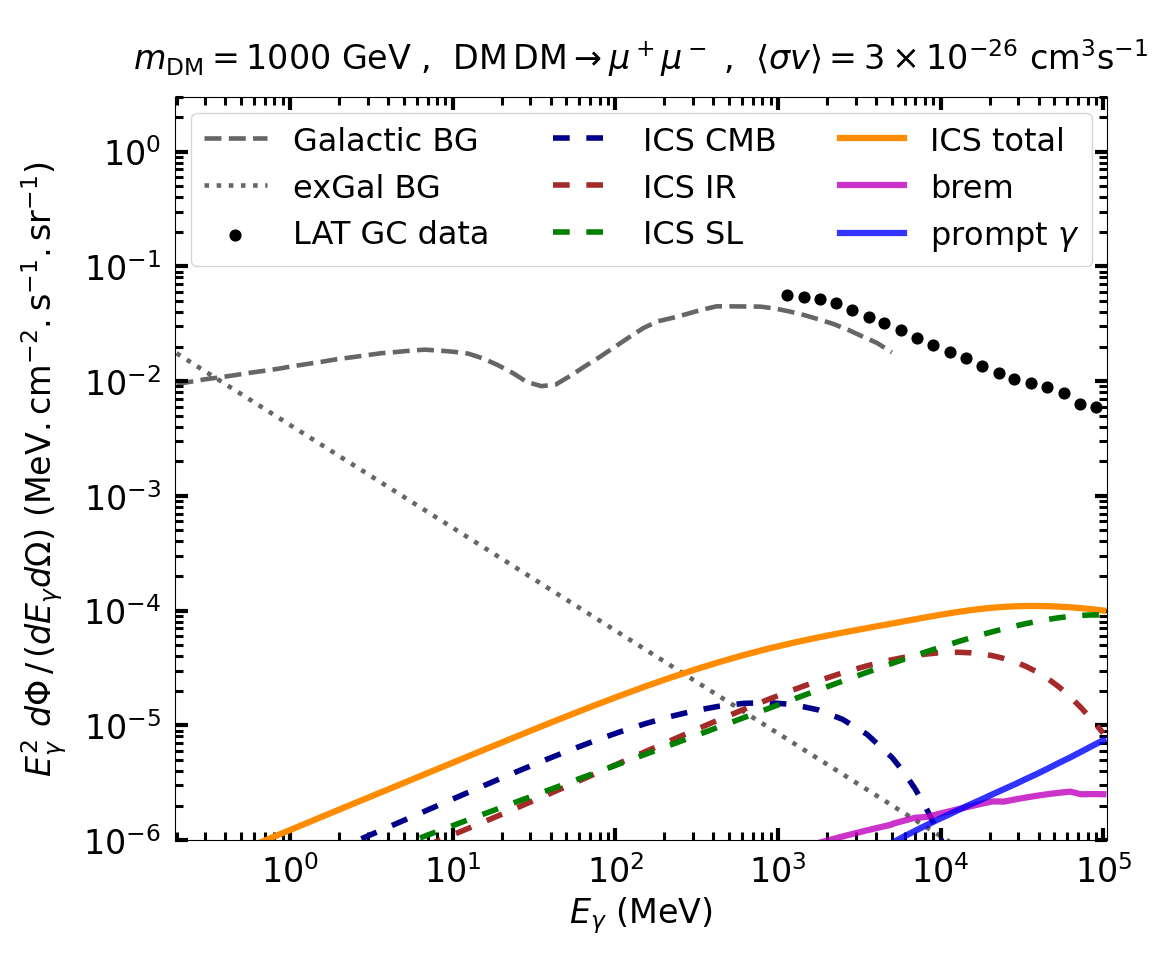}
\vspace{-6mm}
\caption{\em Different types of {\bfseries DM-induced photon fluxes} discussed in 
eqs.~\eqref{eq:prompt_gamma}, \eqref{eq:ICS_flux} and \eqref{eq:brem_flux}, 
averaged over a disk of $10^{\circ}$ around the GC. 
The four panels are for four different values of the DM mass: 
$m_{\rm DM} = 1, 10, 10^2$ and $10^3$ GeV. 
Here, for definiteness, we consider ${\rm DM \, DM} \rightarrow \mu^+\mu^-$ annihilations with 
$\langle \sigma v \rangle = 3\times 10^{-26}$ $\rm cm^3 s^{-1}$. 
In addition, different diffuse photon backgrounds and 
the {\sc Fermi-Lat} GC $\gamma$-ray data are also shown (see the text for details). 
This figure shows that, in our energy range of interest, 
the secondary photon flux such as the ICS can be a very important component of the DM signal for different DM masses.}
\label{fig:DM_fluxes}
\end{figure*}

\smallskip 
In fig.~\ref{fig:DM_fluxes}, the blue curves show the prompt $\gamma$-ray fluxes discussed 
in eq.~\eqref{eq:prompt_gamma} (averaged over the observation region $\Delta \Omega$), 
for four benchmark values of the DM mass: 
$m_{\rm DM} = 1, 10, 10^2$ and $10^3$ GeV, 
considering the NFW DM profile (eq.~\ref{eq:rho_profile}) and the 
${\rm DM \, DM} \rightarrow \mu^+\mu^-$ annihilation channel 
with $\langle \sigma v \rangle$ set at the thermal value 
$\langle \sigma v \rangle = 3\times 10^{-26}$ $\rm cm^3 s^{-1}$.

\subsubsection{Secondary signal: Inverse Compton Scatterings}
\label{sec:secondary_ICS}

We now move to consider the photon fluxes generated due to the ICS of 
DM-induced $e^\pm$ off the ambient photon bath, 
composed of mainly CMB, infrared (IR) and starlight (SL), 
collectively denoted as the Inter-Stellar Radiation Field (ISRF). 
The $e^-/e^+$ population produced by DM annihilations is written in terms of the source function as:
\begin{equation}
Q_e (E^S_e, r) = \frac{\langle \sigma v \rangle}{2 \, m^2_{\rm DM}} \, \frac{dN_e}{dE^S_e} \, \rho^2_{\rm DM}(r) \, ,
\end{equation}
where $\frac{dN_e}{dE^S_e} (E^S_e, m_{\rm DM})$ is the energy spectrum of $e^-/e^+$ sourced by DM annihilation into a given channel. 
Such spectra are estimated using the analytical expressions in \cite{Cirelli:2020bpc} for $m_{\rm DM} < 5$ GeV and 
using the tools in \cite{PPPC_web} for $m_{\rm DM} \ge 5$ GeV. 

The $e^\pm$, after being produced from DM annihilation, 
propagate through the galactic medium undergoing various effects, such as spatial diffusion, advection, convection, re-acceleration and
radiative energy losses, and give rise to a steady state distribution \cite{Evoli:2016xgn, Evoli:2017vim, DelaTorreLuque:2023olp}. 
Since we will be interested in a region of the galactic halo that 
does not cover the scale of the accretion region of the central black hole, 
we will assume that the effects of advection and convection can be neglected.
On the other hand, near the region around the GC, 
for $e^\pm$ with energies above a GeV the effect of 
energy loss via various radiative processes becomes 
dominating over other processes, 
e.g.~spatial diffusion and advection \cite{Bartels:2017dpb}. 
Also, many of the above-mentioned processes turn out to be 
more relevant for the propagation of the cosmic-ray 
nuclei rather than for the $e^\pm$ (see, e.g., \cite{Delahaye:2008ua, Evoli:2016xgn} or \cite{Cirelli:2024ssz} for a summary).
We will therefore adopt here a simplified treatment by assuming that 
energy losses via various radiative processes dominate 
the propagation of the DM-induced $e^\pm$. 
This assumption, referred to as the `on the spot' approximation, 
gives a semi-analytic way to solve the propagation of $e^\pm$ in the Galaxy. 
However, {in Sec.~\ref{sec:propagation}} we will illustrate the effects 
caused by considering the full propagation of $e^\pm$,  
and see that the corresponding DM signals from our target region 
{(and hence the limits on DM based on that)} 
vary at most by a factor of few for the DM mass in 
the usual GeV--TeV range. 
A somewhat similar result was previously obtained in \cite{DiMauro:2021qcf} and \cite{Meade:2009iu, Cirelli:2009vg, Bartels:2017dpb}. 

With the above-mentioned assumption, the spatial and energy 
distribution of the steady state $e^\pm$'s that give rise to 
the ICS flux is estimated as \cite{Cirelli:2023tnx, Cirelli:2020bpc, Cirelli:2009vg}:
\begin{equation}
\frac{dn_e}{dE_e} (E_e, \Vec{x}) = \frac{1}{b_{\rm tot} (E_e, \Vec{x})} 
\int^{m_{\rm DM}}_{E_e} dE^S_e \, Q_e (E^S_e, r) \, ,
\label{eq:dnedE}
\end{equation}
where $b_{\rm tot} (E_e, \Vec{x})$ is the total energy loss rate of the 
$e^\pm$ (having an energy $E_e$) at a position $\Vec{x}$ in the Galaxy. 
This $b_{\rm tot} (E_e, \Vec{x})$ includes energy losses of $e^\pm$ 
by various radiative processes, i.e.~ICS off the ambient photon baths, 
synchrotron emission in the Galactic magnetic field, 
Coulomb interactions with the interstellar gases, ionization of the same gases 
and bremsstrahlung on the same gases. 
We refer the reader to \cite{Buch:2015iya} for their detailed expressions and just remind here that the energy losses through the Coulomb interaction, ionization and bremsstrahlung 
are more important when the $e^\pm$ energy $E_e$ is 
below a GeV, while for $E_e$ above a GeV, the losses due to ICS and 
synchrotron dominate. 

Using the $e^+/e^-$ distribution obtained in eq.~\eqref{eq:dnedE} one can estimate the total ICS emissivity as: 
\begin{equation}
j_{\rm ICS} (E_\gamma, \Vec{x}(s,b,l)) = 2 \, \int^{m_{\rm DM}}_{m_e} dE_e \, 
\sum_{i\in {\rm ISRF}} \hspace{-2mm} \mathcal{P}^i_{\rm ICS} (E_\gamma, E_e, \Vec{x}) 
\,\, \frac{dn_e}{dE_e}(E_e,\Vec{x}) \, ,
\label{eq:j_ICS}
\end{equation}
where the factor of 2 takes into account the contributions of both electrons and positrons. The coordinates $(s,b,l)$ correspond to respectively the $l.o.s.$ linear coordinate, latitude and longitude of a point with respect to the observer. 
The angular distance $\theta$ from the GC is given by $\cos \, \theta = \cos \, b \, \cos \, l$, and the vertical distance above the galactic 
plane $z = s \, \sin\,b$.  
The functions $\mathcal{P}^i_{\rm ICS}$ denote the differential power emitted into photons with energy $E_\gamma$ 
due to the IC scatterings of an $e^+/e^-$ (having energy $E_e$) 
on each ambient photon bath $i$ in the ISRF (see \cite{Cirelli:2009vg, Colafrancesco:2005ji, Cirelli:2024ssz} and Sec.~\ref{sec:appendix_ISRF_model}), 
\begin{equation}
\mathcal{P}^i_{\rm ICS} (E_\gamma, E_e, \Vec{x}) = c \, E_\gamma \int d\epsilon \, n^{\rm ISRF}_i (\epsilon, \Vec{x}) \,\, 
\sigma_{\rm IC} (\epsilon, E_\gamma, E_e) \, ,
\label{eq:P_ICS}
\end{equation}
where $i\rightarrow$ CMB, IR or SL. 
{The differential photon number density $n^{\rm ISRF}_i (\epsilon, \Vec{x})$ corresponding to 
each ISRF component is the same one used in \cite{Cirelli:2020bpc}.} 
The quantity $\sigma_{\rm IC}$ is the Klein-Nishina cross-section, 
obtained using the analytical expressions given in 
\cite{Cirelli:2009vg, Colafrancesco:2005ji, Cirelli:2024ssz}. 
The limit of integration over $\epsilon$ in eq.~\eqref{eq:P_ICS} is 
determined by the  kinematics of the IC scattering 
$1 \le q \le m^2_e/4E^2_e$, where 
$q \equiv \frac{E_\gamma m^2_e}{4 \epsilon E_e (E_e-E_\gamma)}$. 
An illustration of $P^i_{\rm ICS}$ (corresponding to the three ISRF components) 
as a function of $E_\gamma$ for different input $e^\pm$ energies can be 
found in figure 1 of \cite{Cirelli:2020bpc}.

Finally, the observable ICS photon flux 
(averaged over the observation region $\Delta \Omega$) is given by 
\cite{Cirelli:2020bpc, Cirelli:2023tnx}:
\begin{equation}
\frac{d\Phi_{\rm ICS}}{dE_\gamma d\Omega} = \frac{1}{\Delta \Omega} \int_{\Delta \Omega} d\Omega \, 
\left[\frac{1}{E_\gamma} \int_{l.o.s.} ds \,\, \frac{j_{\rm ICS} (E_\gamma, \Vec{x}(s,b,l))}{4\pi} \right] \, .
\label{eq:ICS_flux}
\end{equation}
For the spatial integral in the above equation, we cut the distribution of the $e^\pm$ density 
in the radial direction at 
$R = R_{\rm Gal} = 20 \, {\rm kpc}$, and in the vertical direction  
at $z = L_{\rm Gal} = 4 \, {\rm kpc}$, 
assuming this to be the size of the zone that keeps the $e^\pm$ confined.

\medskip

In fig.~\ref{fig:DM_fluxes} we plot the ICS photon fluxes 
(averaged over the observation region), 
for four values of the DM mass: $m_{\rm DM} = 1, 10, 10^2$ and $10^3$ GeV, 
considering the ${\rm DM \, DM} \rightarrow \mu^+\mu^-$ annihilation channel with 
$\langle \sigma v \rangle = 3\times 10^{-26}$ $\rm cm^3 s^{-1}$. 
We show contributions from different target photons (i.e., from CMB, IR, and SL) separately, 
together with the sum of all these contributions (the orange solid lines). 
Fig.~\ref{fig:DM_fluxes} shows that, for different DM masses, the DM-induced ICS flux in principle can dominate 
over the other types of DM-induced signals in our energy range of interest, especially in the low energy regime.

\subsubsection{Secondary signal: bremsstrahlung} 
\label{sec:secondary_brem}

The same $e^\pm$ populations that produce the ICS flux 
can also give rise to the bremsstrahlung emission.  
In analogy with eq.~\eqref{eq:j_ICS}, the bremsstrahlung emissivity 
can be expressed as: 
\begin{equation}
j_{\rm brem} (E_\gamma, \Vec{x}(s,b,l)) = 2 \, \int^{m_{\rm DM}}_{m_e} dE_e \, 
\mathcal{P}_{\rm brem} (E_\gamma, E_e, \Vec{x}) \,\, \frac{dn_e}{dE_e}(E_e,\Vec{x}) \, ,
\end{equation}
where the $e^\pm$ distribution $\frac{dn_e}{dE_e}(E_e,\Vec{x})$ is given 
by eq.~\eqref{eq:dnedE}. The bremsstrahlung power emitted 
into photons with energy $E_\gamma$ due to the scattering of an $e^\pm$ 
of energy $E_e$ (with $E_e > E_\gamma$) is given by \cite{Buch:2015iya}: 
\begin{equation}
\mathcal{P}_{\rm brem} (E_\gamma, E_e, \Vec{x}) = c \, E_\gamma \sum_{i} 
n_i(\Vec{x}) \, \frac{d\sigma^{\rm brem}_i}{dE_\gamma} (E_e, E_\gamma) \, .
\end{equation}
Here $n_i(\Vec{x})$ describes the number density distribution of each of the 
gas species (ionic, atomic and molecular) and 
$\frac{d\sigma^{\rm brem}_i}{dE_\gamma} (E_e, E_\gamma)$ is the 
corresponding differential scattering cross-section for bremsstrahlung. 
For all the details of these expressions, we refer again the reader to \cite{Buch:2015iya}. 

\medskip

Finally, the observable photon flux 
 due to bremsstrahlung (averaged over the observation region $\Delta \Omega$) is computed as: 
\begin{equation}
\frac{d\Phi_{\rm brem}}{dE_\gamma d\Omega} = \frac{1}{\Delta \Omega} 
\int_{\Delta \Omega} d\Omega \, \left[\frac{1}{E_\gamma} \int_{l.o.s.} ds \,\, 
\frac{j_{\rm brem} (E_\gamma, \Vec{x}(s,b,l))}{4\pi} \right] \, .
\label{eq:brem_flux}
\end{equation}

The photon fluxes due to the bremsstrahlung emissions arising for different DM masses are shown 
in fig.~\ref{fig:DM_fluxes} 
by the magenta curves. For our target region, bremsstrahlung emissions 
can dominate the DM signal in the intermediate part of our energy 
range of interest, especially when the DM mass is relatively smaller.

\subsection{Background models and data}
\label{sec:background}

The total diffuse photon background towards our target region 
(i.e., the $10^{\circ}$ cone around the GC) 
receives contributions from both galactic and extra-galactic origins.

The galactic photon backgrounds are obtained from \cite{ODonnell:2024aaw} for the energy range 
$0.2 \, {\rm MeV} \lesssim E_\gamma \\ \lesssim 5 \, {\rm GeV}$. 
This is the energy range that we use for the study of future MeV telescopes. 
The diffuse galactic background model consists of mainly four different astrophysical components. Three of them, namely:  
a bremsstrahlung component, a $\pi^0$ component and a high energy ICS component (${\rm ICS}_{\rm hi}$), 
were computed with the \texttt{GALPROP} code \cite{Strong:1998pw} and fitted to the {\sc Fermi-Lat} data 
at higher energies \cite{2011crpa.conf..473S}. 
The fourth one, another ICS component (${\rm ICS}_{\rm lo}$), was modeled as a power-law and 
fitted to {\sc Comptel} and {\sc Egret} data at lower energies \cite{Beacom:2005qv, Bartels:2017dpb}.  
All these background models were presented in \cite{Bartels:2017dpb} for a region 
$|l| \leq 5^\circ$, $|b| \leq 5^\circ$ around the GC. 
Each of these models was then rescaled by \cite{ODonnell:2024aaw} to the 
$10^{\circ}$ cone region around the GC and presented in their figure 1. 
We adopt these models as our fiducial galactic photon backgrounds. 
On the other hand, for the extra-galactic component, we adopt the single power law model from \cite{Ray:2021mxu}, 
with their best-fit values of the normalization and the power law index as the fiducial parameters. 

\medskip

In each panel of fig.~\ref{fig:DM_fluxes}, the total 
fiducial galactic photon background 
(the sum of four components listed above) and the fiducial extra-galactic photon background 
are shown by the gray dashed and dotted lines, respectively.  
Along with these backgrounds, we also show in each plot with black points the 
{\sc Fermi-Lat} $\gamma$-ray data towards the GC. 
These are taken from \cite{Fermi-LAT:2015sau} 
for a $15^{\circ}\times15^{\circ}$ 
region around the GC and presented in this figure by normalizing with respect to the corresponding solid angle.

\section{Future MeV telescopes}\label{sec:future_telescopes}

We study the prospects of probing the DM signals in 
different near-future MeV space telescopes~\cite{Engel:2022bgx}, such as 
{\sc Amego}, {\sc e-Astrogam} and {\sc Mast}. 
Apart from these three, there are several other planned 
or proposed space-based MeV telescopes: for instance,  
{\sc Cosi}~\cite{Caputo:2022dkz}, {\sc Gecco}~\cite{Orlando:2021get}, 
{\sc Adept}~\cite{Hunter:2013wla}, {\sc Grams}~\cite{Aramaki:2021o5} 
and {\sc Pangu}~\cite{Wu:2014tya}. These instruments have 
effective areas which are either smaller or of the same order 
of magnitude than the three above-mentioned instruments, and therefore are expected to provide comparatively 
smaller or similar sensitivities for the DM signal searches. 
We thus choose to focus on {\sc Amego}, 
{\sc e-Astrogam} and {\sc Mast}, and give here a brief description of their main properties relevant for DM searches.

\begin{itemize}

\item[$\circ$] \underline{\sc Amego}: 
The All-sky Medium Energy Gamma-ray Observatory ({\sc Amego}) is a proposed future space-based mission, poised to provide important contributions to multi-messenger astrophysics in the late 2020's and beyond. 
It can operate in two different modes, Compton scattering and pair-production modes, 
to achieve high sensitivity in a wide energy range 0.2 MeV to $\sim5$ GeV. 
The Compton mode is divided into two parts: untracked Compton and tracked Compton. 
By combining three different event classifications it can achieve an effective area of 
$\sim 500 - 1000 \, {\rm cm^2}$ across the full energy range.  
A detailed overview of various instrumental details of {\sc Amego}, e.g., effective area, 
energy resolution and angular resolution for different modes, 
can be found in Refs. \cite{AMEGO:2019gny, 2020SPIE11444E..31K}. 
For our work we use the three effective areas discussed above 
and an energy resolution of 
30\% for the full range of energy over which it can operate. 
There is also a more recent proposal for an instrument, named 
{\sc Amego-X}~\cite{Caputo:2022xpx}, with a similar concept.

\item[$\circ$] \underline{\sc e-Astrogam}: {\sc e-Astrogam} 
(or enhanced-{\sc Astrogam}) 
is a satellite 
gamma-ray mission concept proposed by a wide 
international 
community for the late 2020's. Like {\sc Amego}, 
it can operate in both Compton scattering 
and pair-production modes to 
achieve good sensitivity over a broad energy range 
0.3 MeV to 3 GeV 
(the lower energy limit can be pushed to energies as low as 
150 keV and 30 keV with improved tracker and calorimetric 
detections, respectively). 
Its effective area and energy resolution vary between 
$\sim10^2 - 10^3 \, {\rm cm^2}$ 
and $\mathcal{O}(1)\% - 30\%$, respectively, in the two modes 
over the energy range 0.3 MeV - 3 GeV. 
All the details related to its various instrumentation can 
be obtained from 
\cite{e-ASTROGAM:2016bph, e-ASTROGAM:2017pxr}. 
In our case, we conservatively consider an energy 
resolution of 30\% over its full energy range. 

\item[$\circ$] \underline{\sc Mast}: The Massive Argon Space Telescope ({\sc Mast}) is another proposed future 
satellite-based telescope that plans to use a liquid Argon time projection chamber for $\gamma$-ray 
astronomy. Compared to the other future telescopes, it has a significantly larger effective area, which is estimated to be approximately $\sim 10^5 \, {\rm cm^2}$ over the 
energy range $E_\gamma > 10$ MeV. For the instrumental details see \cite{Dzhatdoev:2019kay}. 
We have considered an energy resolution of 30\% for the full energy range of this telescope. 
The high sensitivity of {\sc Mast} in searching MeV DM and primordial black hole (PBH) DM signals was discussed in 
\cite{ODonnell:2024aaw, Coogan:2021sjs, Berlin:2023qco} 
and \cite{Coogan:2020tuf, Auffinger:2022dic}, respectively.

\end{itemize}

\begin{figure*}[ht!]
\centering
\includegraphics[width=0.53\textwidth]{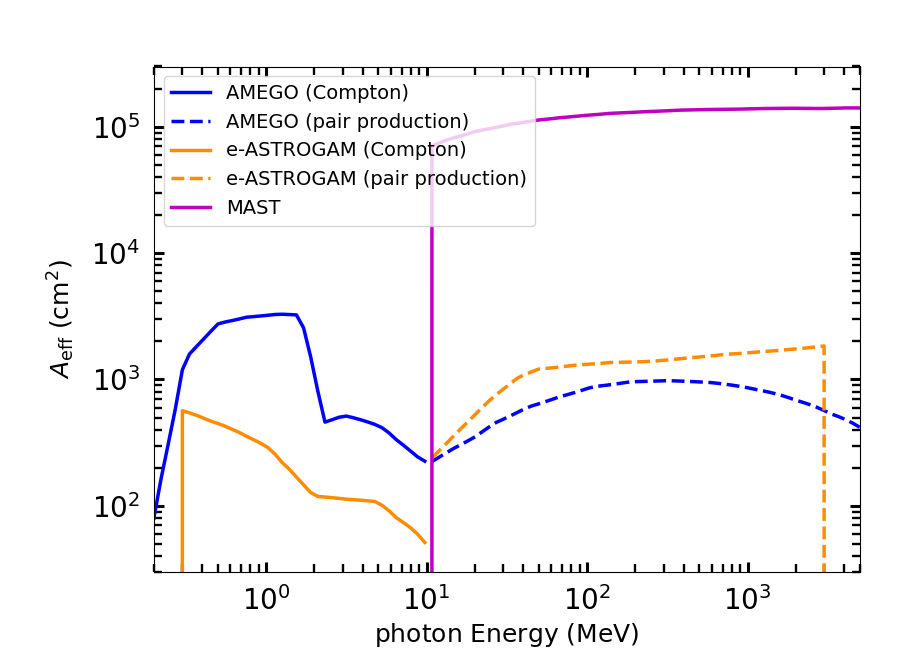}
\vspace{-1mm}
\caption{\em {\bfseries Effective areas} of the future telescopes {\sc Amego}, {\sc e-Astrogam} and {\sc Mast}, for the range of the photon energy considered in this work.}
\label{fig:A_eff}
\end{figure*}

In fig.~\ref{fig:A_eff} the effective areas of the various considered future 
telescopes are shown for the range of the photon energy 
(0.2 MeV - 5 GeV) considered in this work. 
See also \cite{ODonnell:2024aaw} 
for a more detailed summary of these telescopes. 

\section{Methodology}
\label{sec:methodology}
 
In this section we discuss the methodology adopted in this work to obtain  
the projections of the future MeV telescopes on the DM parameter space. 
To obtain these projections we consider the sum of ICS \eqref{eq:ICS_flux}, 
bremsstrahlung \eqref{eq:brem_flux} and 
prompt $\gamma$-rays \eqref{eq:prompt_gamma} fluxes discussed 
in section~\ref{sec:signals} as the total DM-induced signal.

\subsection{Conservative upper bounds}\label{sec:upper_bounds}

Before discussing the projections for the 
future MeV telescopes, we first 
derive some simple estimates of the possible upper bounds on the 
DM annihilation cross-section $\langle \sigma v \rangle$, 
based on existing $\gamma$-ray observations 
(in the range $0.2 \, {\rm MeV} \lesssim 
E_\gamma \lesssim 100 \, {\rm GeV}$).  

For each of the considered annihilation channels,  
we obtain such estimates using the following approach. 
We require that, for a fixed $m_{\rm DM}$, the total 
predicted DM photon signal, averaged over 
the $10^{\circ}$ cone around the GC, 
does not exceed the total fiducial MeV photon background 
(the sum of the gray dashed and dotted lines from 
fig.~\ref{fig:DM_fluxes}, based on {\sc Egret} and {\sc Comptel} data 
as discussed in sec.~\ref{sec:background}) 
at any energy bin in the range 
$0.2 \, {\rm MeV} \lesssim E_\gamma \lesssim 5 \, {\rm GeV}$, 
as well as the total signal, averaged over the 
$15^{\circ}\times15^{\circ}$ region 
around the GC, does not exceed the {\sc Fermi-Lat} data 
(the black points in fig.~\ref{fig:DM_fluxes}) 
at any energy bin in the range 
$5 \, {\rm GeV} \lesssim E_\gamma \lesssim 100 \, {\rm GeV}$.   
This leads to an upper bound on 
$\langle \sigma v \rangle$ for the given $m_{\rm DM}$. 
The resulting constraints are drawn with a red solid line in 
{figures \ref{fig:sv_mx_limits_1} and \ref{fig:sv_mx_limits_2}} below.

\subsection{Discovery reach of future MeV telescopes}

In order to obtain the projected sensitivities 
or the discovery potentials of the 
MeV telescopes we take two approaches, discussed below. 

\subsubsection{Simple signal-to-noise ratio (SNR) method}\label{sec:SNR_method}

In this case we take the simple approach 
presented in \cite{Coogan:2020tuf, Coogan:2021sjs} 
and require the signal-to-noise ratio (SNR) over the observation time period to be larger than five, i.e., 
\begin{equation}
\frac{\left.N_\gamma\right\vert_{\rm DM}}{\sqrt{\left.N_\gamma\right\vert_{\rm BG}}} \, \ge 5 \, ,
\label{eq:SNR_method}
\end{equation} 
which leads to a $\sim5\sigma$ projection on $\langle \sigma v \rangle$ 
for a given $m_{\rm DM}$. Here $\left.N_\gamma\right\vert_{\rm BG}$ is the photon count of the 
total diffuse background (sum of the galactic and extra-galactic background components discussed in sec.~\ref{sec:signals}), 
while $\left.N_\gamma\right\vert_{\rm DM}$ 
corresponds to the DM induced total photon signal for a given DM mass. 
The number of photon count over the energy range 
[$E_{\rm min}, E_{\rm max}$] is defined as: 
\begin{equation}
N_\gamma = t_{\rm obs} \, \int^{E_{\rm max}}_{E_{\rm min}} 
dE_\gamma \, A_{\rm eff} (E_\gamma) \, \int_{\Delta \Omega} 
d\Omega \, \frac{d\Phi}{dE_\gamma d\Omega} \, ,
\label{eq:N_gamma} 
\end{equation}
with 
\begin{equation}
\frac{d\Phi}{dE_\gamma d\Omega} = \int dE'_\gamma \, R_{\epsilon} (E_\gamma, E'_\gamma) \, 
\frac{d\Phi}{dE'_\gamma d\Omega} \, ,
\label{eq:flux_eng_resol}
\end{equation} 
where $\frac{d\Phi}{dE'_\gamma d\Omega}$ corresponds to either 
the DM signal or the diffuse background discussed in sec.~\ref{sec:signals}. 
The function $R_{\epsilon} (E_\gamma, E'_\gamma)$ is a gaussian with mean 
$E'_\gamma$ and standard deviation $\epsilon(E'_\gamma) E'_\gamma$ that accounts for the 
finite energy resolution of the telescope \cite{Boddy:2015efa, Perelstein:2010at}. 
The energy resolution $\epsilon$ and the effective area $A_{\rm eff}$ 
of different MeV telescopes used here are discussed in sec.~\ref{sec:future_telescopes}. 
The solid angle $\Delta \Omega$ corresponds to our ROI, a region of 
$10^{\circ}$ radius from the GC.  
Note that the sensitivity of the projection scales with the 
observation time as $\sqrt{t_{\rm obs}}$.

\subsubsection{Fisher method}\label{sec:Fisher_method}

In this case we take an approach similar to the one presented in \cite{ODonnell:2024aaw, Coogan:2021rez} 
involving the Fisher matrix method, employed previously in \cite{Edwards:2017mnf}. 
We define the vector $\Vec{\theta}$ involving the signal and 
the background parameters:
\begin{equation}
\Vec{\theta} = \left[ \Gamma^{\rm SIG}, \, \theta^{\rm BG}_{\rm brem}, \, 
\theta^{\rm BG}_{\pi^0}, \, \theta^{\rm BG}_{{\rm ICS_{hi}}}, \, 
\theta^{\rm BG}_{{\rm ICS_{lo}}}, \, \theta^{\rm BG}_{\rm e.g.} \right] \, ,
\end{equation}
where $\Gamma^{\rm SIG}$ is the normalization of the DM signal 
(i.e., $\langle \sigma v \rangle$ for a fixed $m_{\rm DM}$ and a fixed annihilation channel), 
while $\theta^{\rm BG}_i$ denotes the rescaling of the normalization for each 
background component $i$ (representing the bremsstrahlung, $\pi^0$, ${\rm ICS_{hi}}$, 
${\rm ICS_{lo}}$ and extra-galactic components) with respect to the fiducial one, discussed 
in sec.~\ref{sec:background}. 
The total differential photon flux is defined as:
\begin{equation}
\phi_{\rm tot} (\Vec{\theta}) = \frac{d\Phi^{\rm SIG}}{dE_\gamma d\Omega} 
(\Gamma^{\rm SIG}) + \sum_{I} \, \theta^{\rm BG}_I \, \left\lbrace \frac{d\Phi^{I}_{\rm BG}}{dE_\gamma d\Omega} \right\rbrace_{\rm fiducial} \, ,
\end{equation}
with $\frac{d\Phi}{dE_\gamma d\Omega}$ defined in eq.~\eqref{eq:flux_eng_resol}. 
Using these we define the Fisher matrix as:
\begin{equation}
\mathcal{F}_{ij} = t_{\rm obs} \, \int^{E_{\rm max}}_{E_{\rm min}} dE_\gamma \, A_{\rm eff} (E_\gamma) \, \int_{\Delta \Omega} d\Omega \,\, 
\left( \frac{1}{\phi_{\rm tot}} \frac{\partial \phi_{\rm tot}}{\partial \theta_i} 
\frac{\partial \phi_{\rm tot}}{\partial \theta_j} \right)_{\Vec{\theta} = \Vec{\theta}_{\rm fiducial}} \, ,
\end{equation}
where $\Vec{\theta}_{\rm fiducial}$ denotes the fiducial parameters 
with $(\Gamma^{\rm SIG})_{\rm fiducial}$ set to zero, i.e.,  
considering the null hypothesis (no DM signal is present in the data) to be true. 
$\mathcal{F}$ here is a $6\times6$ symmetric matrix.  
The $2 \sigma$ projected upper bound on the signal normalization $\Gamma^{\rm SIG}$ is 
then defined as: 
\begin{equation}
\Gamma_{\rm proj}^{\rm SIG} = 2 \, \sqrt{(\mathcal{F}^{-1})_{11}} \, \, .
\end{equation}
Here, too, the sensitivity of the projection scales as $\sqrt{t_{\rm obs}}$.

\section{Results and discussion}
\label{sec:results}

\begin{figure*}[ht!]
\includegraphics[width=0.5\textwidth]{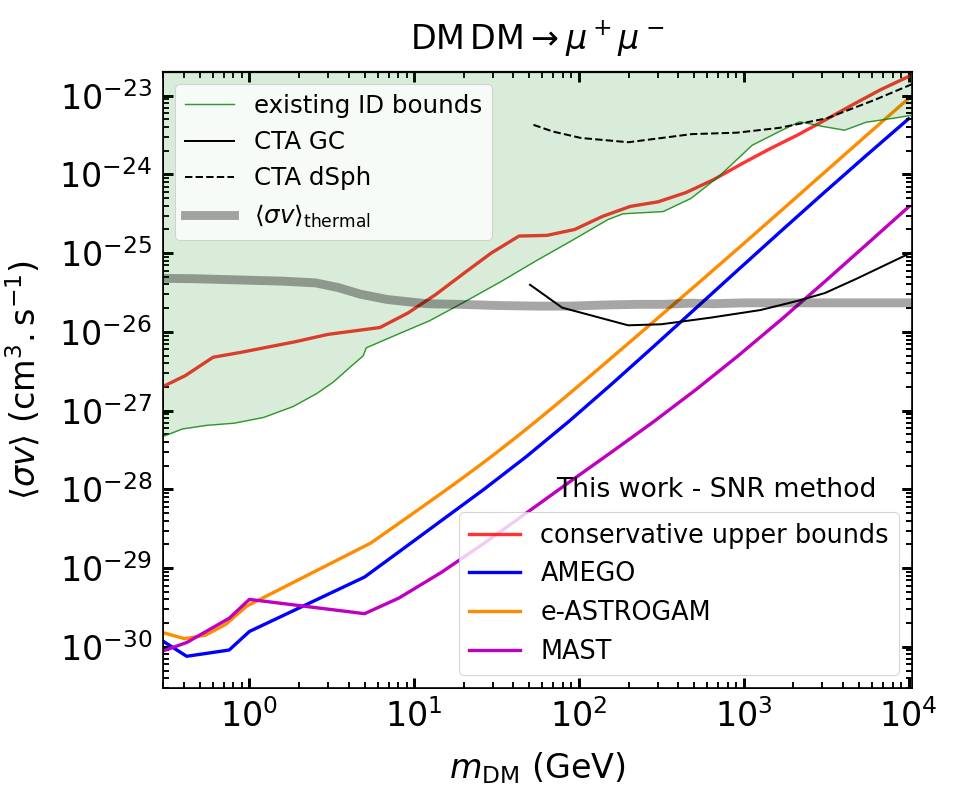}
\includegraphics[width=0.5\textwidth]{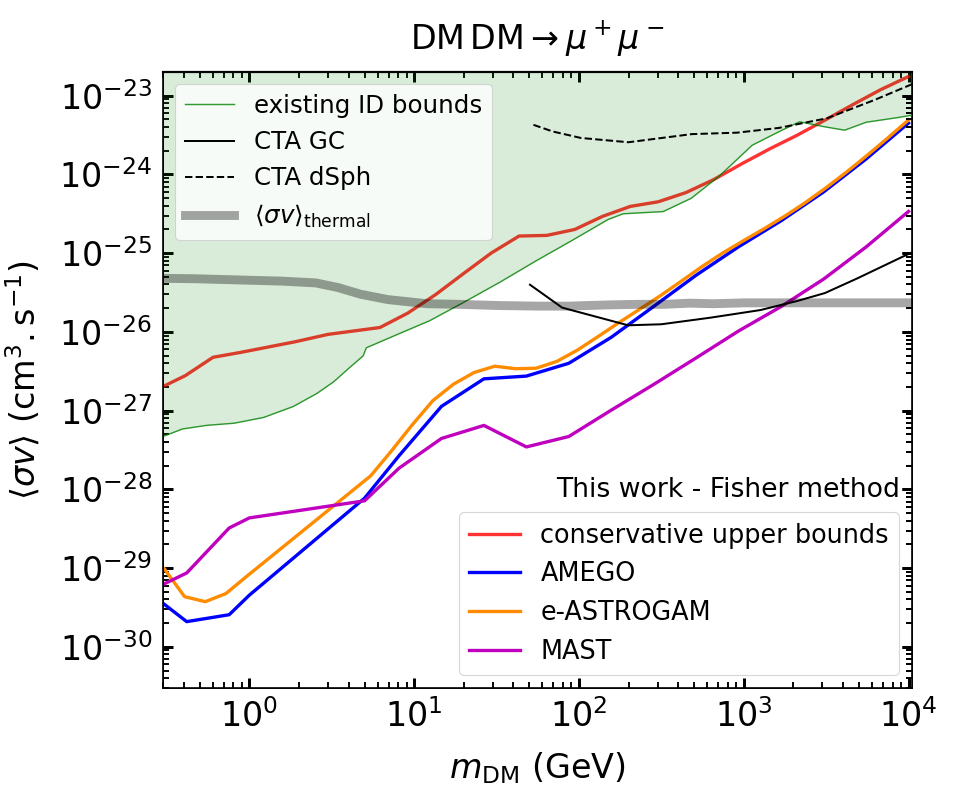} \\
\includegraphics[width=0.5\textwidth]{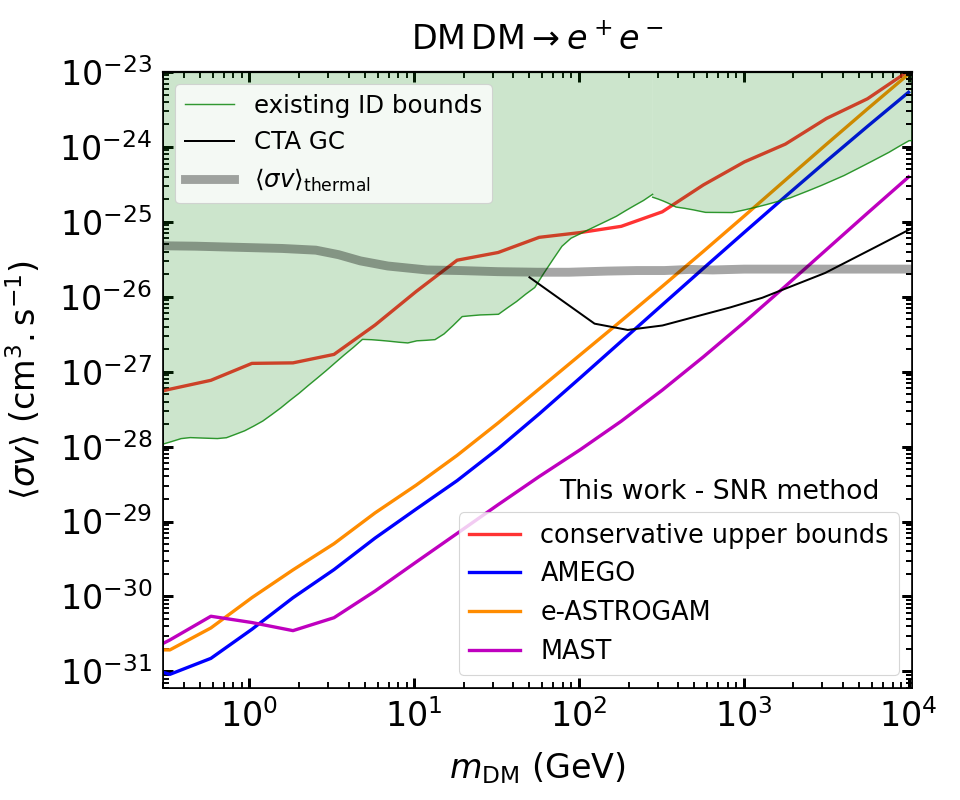}
\includegraphics[width=0.5\textwidth]{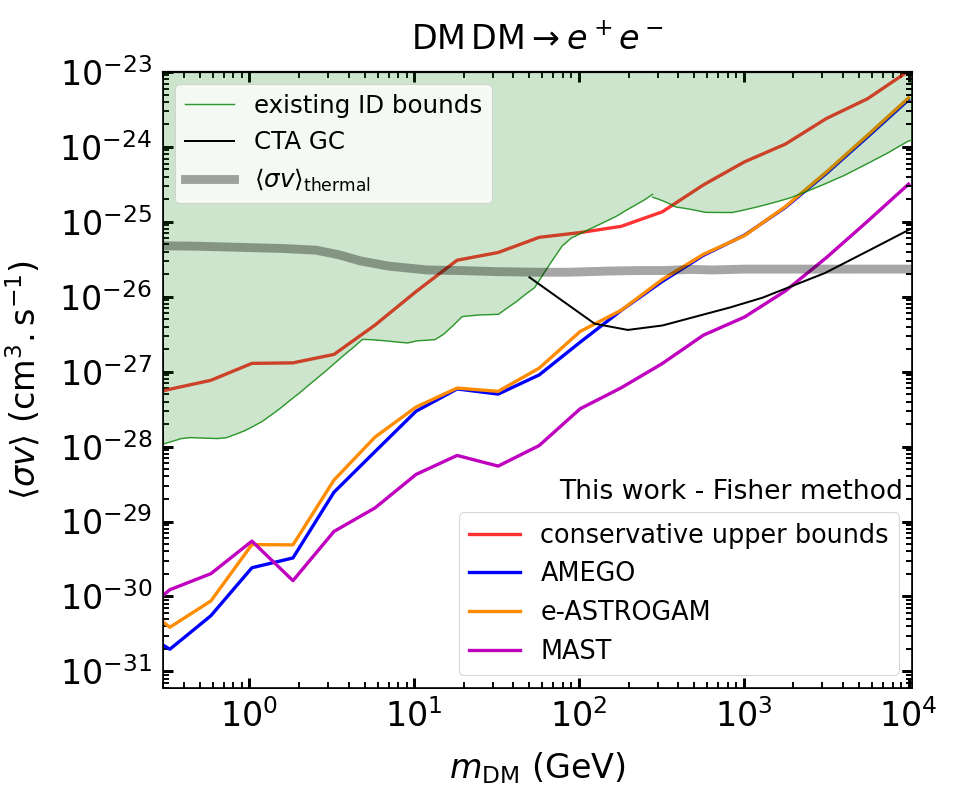}
\vspace{-5mm}
\caption{{\em {\bfseries Upper bounds and projected 
sensitivities} on the annihilation cross-section $\langle \sigma v \rangle$ 
as a function of $m_{\rm DM}$, for $\mu^+\mu^-$ (upper panels) and $e^+e^-$ (lower panels)  
annihilation channels. Notice the different vertical ranges for different channels. 
For each channel, the projected sensitivities are shown considering 
the two statistical approaches discussed in sections~\ref{sec:SNR_method} (left column) 
and \ref{sec:Fisher_method} (right column). 
In each case, our conservative upper bounds are shown with a red solid line, 
and the existing bounds from the literature with green shaded areas (see the text for details). 
The projected sensitivities of the upcoming MeV telescopes {\sc Amego}, 
{\sc e-Astrogam} and {\sc Mast} are shown by blue, orange and magenta curves, respectively, 
for an observation time of $10^8$ sec ($\simeq 3$ yrs).  
The solid and dashed black curves show the projections of {\sc Cta} for 
the observations towards GC and dwarf galaxies (when available), respectively.  
Finally, the solid gray thin band 
in each plot indicates the value of $\langle \sigma v \rangle$ 
corresponding to the observed relic abundance for thermal DM, assuming 
$s$-wave annihilation~\cite{Steigman:2012nb}.}}
\label{fig:sv_mx_limits_1}
\end{figure*}

\begin{figure*}[ht!]
\includegraphics[width=0.5\textwidth]{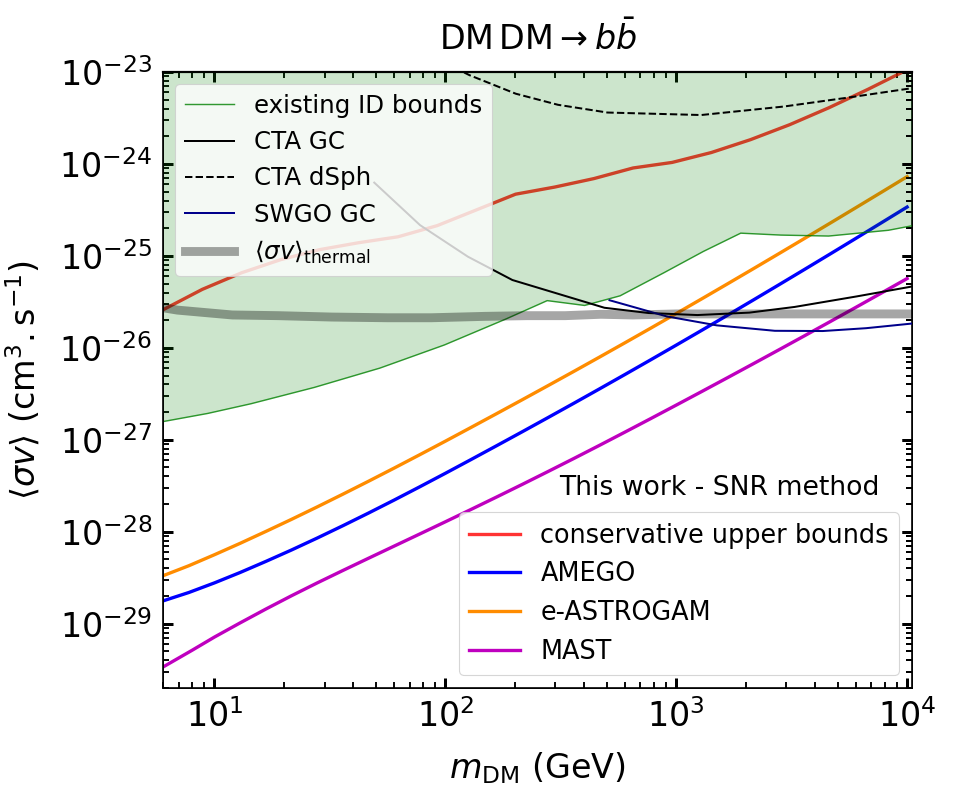}
\includegraphics[width=0.5\textwidth]{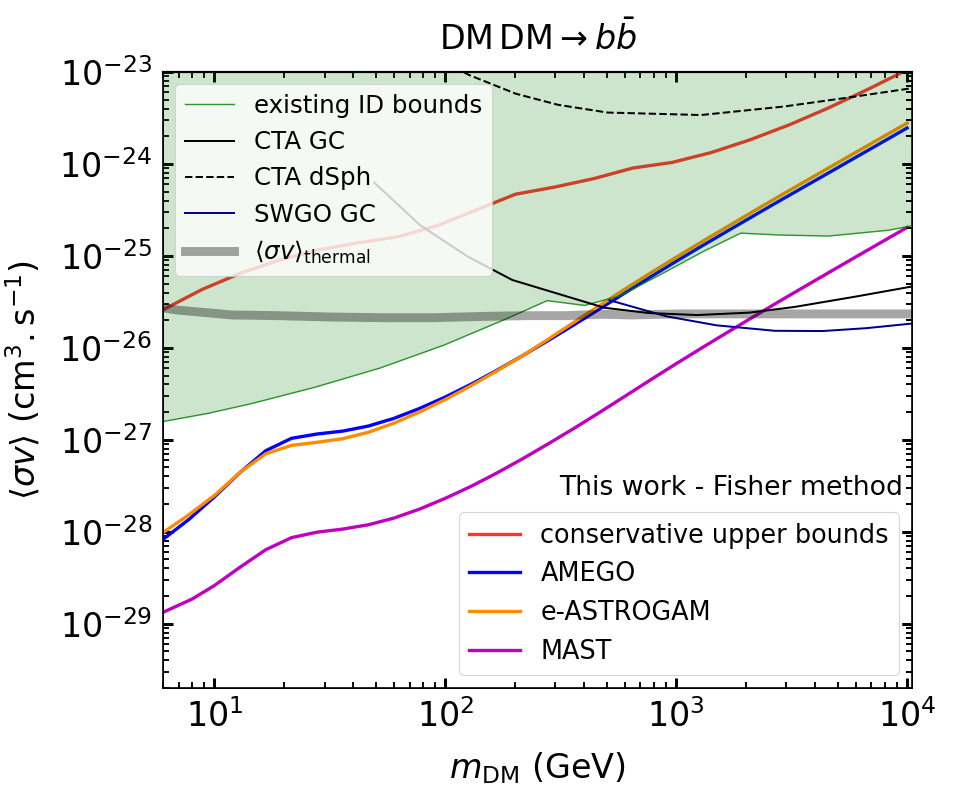} \\
\includegraphics[width=0.5\textwidth]{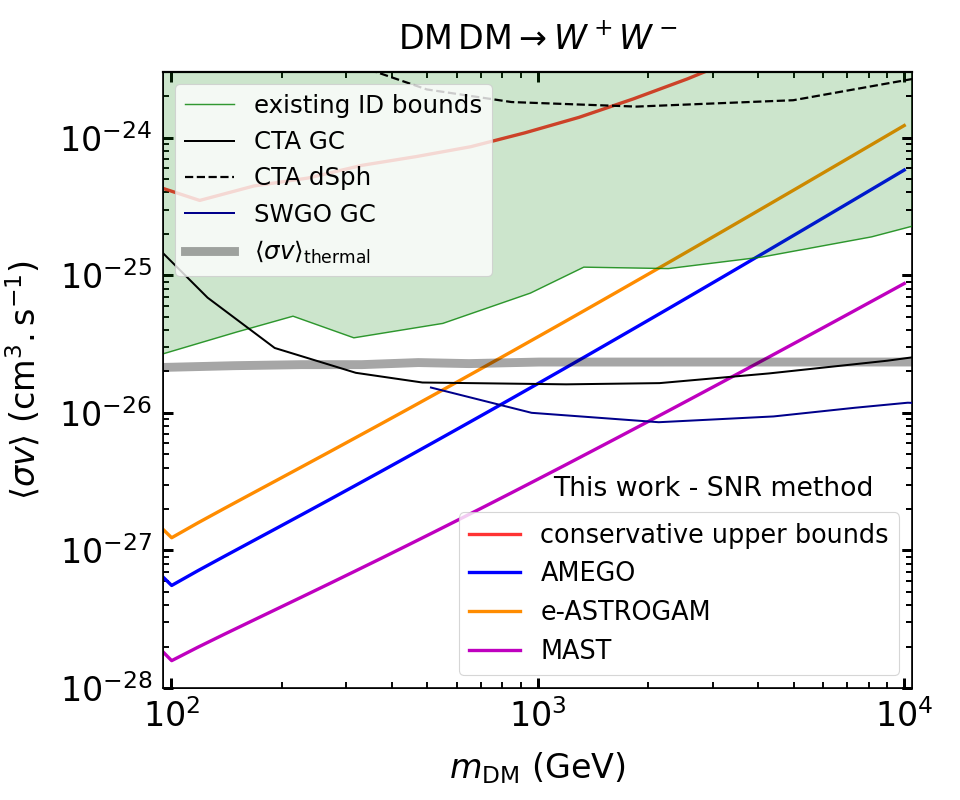}
\includegraphics[width=0.5\textwidth]{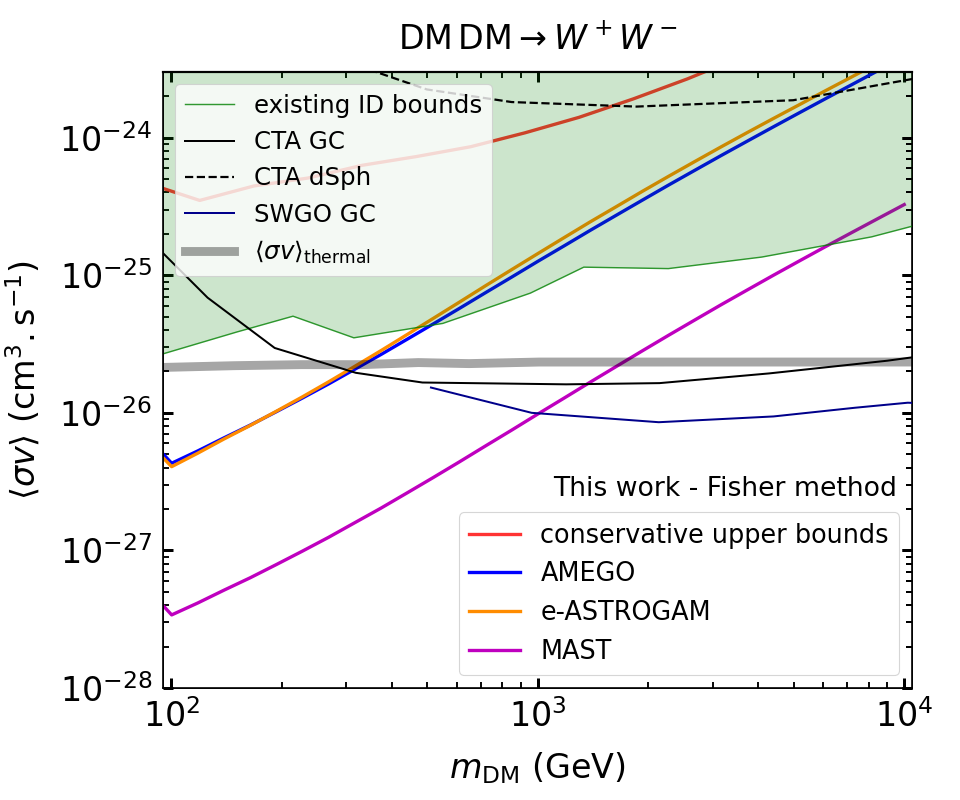}
\vspace{-5mm}
\caption{{\em Same as fig.~\ref{fig:sv_mx_limits_1}, but considering 
$b\bar{b}$ (upper panels) and $W^+W^-$ (lower panels) annihilation channels. 
Notice the different horizontal and vertical ranges for different channels. 
The solid and dashed black curves show the projections of {\sc Cta} for 
the observations towards GC and dwarf galaxies, respectively, 
and the dark blue curve shows the projection for {\sc Swgo}, 
when available.}}
\label{fig:sv_mx_limits_2}
\end{figure*}

In this section we present the bounds and the projected sensitivities of the 
future MeV telescopes on DM annihilation, derived in this work using 
the methodology outlined in sec.~\ref{sec:methodology}. 
The main results of this analysis are presented in 
{fig.~\ref{fig:sv_mx_limits_1} (for $\mu^+\mu^-$ and $e^+e^-$ channels) 
and fig.~\ref{fig:sv_mx_limits_2} (for $b\bar{b}$ and $W^+W^-$ channels)}
for the usual weak-scale DM mass range spanning the GeV--TeV scale. 
We recall that our target of observation is the inner galactic region (see sec.~\ref{sec:signals}) 
from where the secondary emission components such as the ICS flux are expected to be very high 
due to the high densities of the ISRF photons (especially the star light (SL)). 
This helps to increase the reach of the MeV 
telescopes for GeV--TeV DM. 
The high DM density that one expects in this region 
also helps. The different astrophysical ingredients used here are 
the ones discussed in sec.~\ref{sec:signals}. In appendix \ref{sec:astrophysical_models} 
we discuss the variations of such astrophysical ingredients. 

The solid red lines in {figs.~\ref{fig:sv_mx_limits_1} and \ref{fig:sv_mx_limits_2}} 
show our estimated upper bounds on $\langle \sigma v \rangle$ 
as a function of $m_{\rm DM}$, obtained in sec.~\ref{sec:upper_bounds} 
and corresponding to existing MeV-GeV $\gamma$-ray observations, 
essentially from {\sc Comptel}, {\sc Egret} and {\sc Fermi-Lat}.  

Then, for each of the considered DM annihilation channels,  
we present the projected sensitivities of the upcoming MeV telescopes 
{\sc Amego}, {\sc e-Astrogam} and {\sc Mast} with blue, orange and magenta curves, respectively. 
{For each channel in figs.~\ref{fig:sv_mx_limits_1} and \ref{fig:sv_mx_limits_2}, 
the left and right panels correspond to the  
projections for different MeV telescopes} estimated with the two statistical approaches 
discussed in sections~\ref{sec:SNR_method} and \ref{sec:Fisher_method}, respectively. 
The observation time $t_{\rm obs}$ is assumed to be 
$10^8$ sec ($\simeq 3$ yrs) \cite{Coogan:2020tuf, Coogan:2021sjs} 
which is a standard duration achievable by the future MeV telescopes 
considered here~\cite{AMEGO:2019gny, 2020SPIE11444E..31K, Caputo:2022xpx, e-ASTROGAM:2016bph, e-ASTROGAM:2017pxr, Dzhatdoev:2019kay}. 
As pointed out earlier, the projected sensitivities scale with 
the observation time as $\sqrt{t_{\rm obs}}$. 

\begin{figure*}[ht!]
\includegraphics[width=0.5\textwidth]{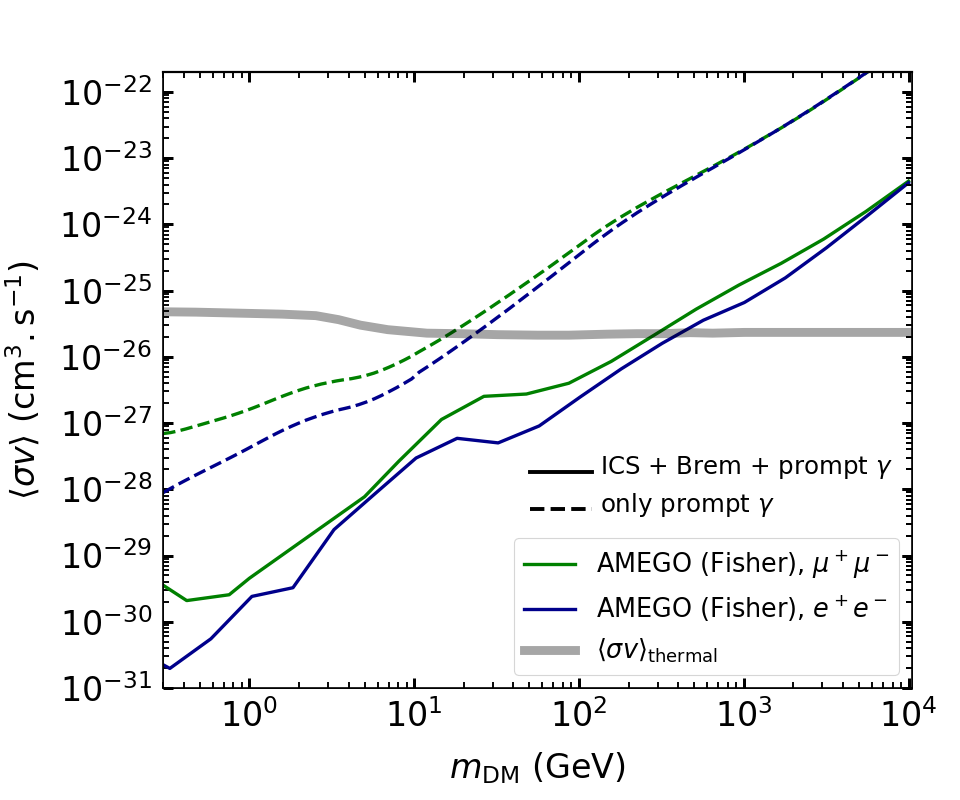}
\includegraphics[width=0.5\textwidth]{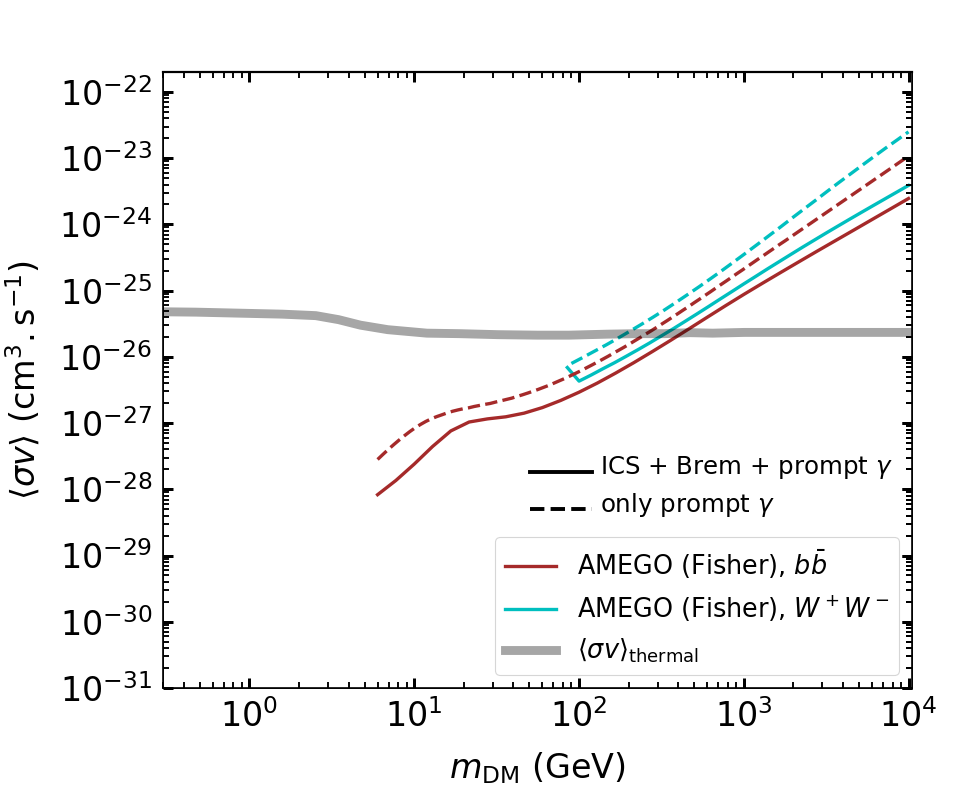}
\vspace{-5mm}
\caption{\em  Illustration of the {\bfseries importance of considering the secondary signals} 
of DM in the case of the leptonic annihilation channels. 
The comparison between the Fisher projections 
(for {\sc Amego} with $t_{\rm obs} = 10^8$ sec) 
obtained considering and not considering the secondary 
signals are shown 
for the $\mu^+\mu^-$, $e^+e^-$ (left panel) and the $b\bar{b}$, 
$W^+W^-$ (right panel) channels. 
The colored solid curves are the projections obtained considering 
the secondary signals and are the same as the ones presented 
in {figs.~\ref{fig:sv_mx_limits_1} and \ref{fig:sv_mx_limits_2}}, 
while the dashed curves show the projections obtained not 
considering the secondary signals.}
\label{fig:sv_mx_limits_diff_components}
\end{figure*}

We compare our results with the existing bounds 
(shown by the green shaded areas in {figs.~\ref{fig:sv_mx_limits_1} and \ref{fig:sv_mx_limits_2}}) 
obtained from the literature. 
For a given annihilation channel, these bounds correspond to the most stringent 
upper limit on $\langle \sigma v \rangle$ (as a function of $m_{\rm DM}$) 
that is compatible with all existing experimental observations. 
The existing bounds for the $\mu^+\mu^-$ (for $m_{\rm DM} > 5$ GeV), 
$b\bar{b}$ and $W^+W^-$ channels are taken from \cite{Cirelli:2024ssz}. 
For the $\mu^+\mu^-$ channel, this bound is a combination 
of limits derived from CMB (assuming $s$-wave annihilation)~\cite{Slatyer:2015jla, Lopez-Honorez:2013cua}, 
dwarf galaxies $\gamma$-rays~\cite{MAGIC:2016xys} and 
{\sc Antares} neutrino~\cite{Albert:2016emp} observations. 
On the other hand, for the $b\bar{b}$ and $W^+W^-$ channels these bounds 
are the convolution of limits obtained from dwarf gamma-rays~\cite{Hess:2021cdp}, 
{\sc Ams} anti-proton~\cite{Calore:2022stf} and 
{\sc Hess} GC~\cite{HESS:2016mib, HESS:2022ygk} observations. 
The bounds for $\mu^+\mu^-$ and $e^+e^-$ for $m_{\rm DM} < 5$ GeV 
are the combinations of limits from {\sc Xmm-Newton} $X$-ray~\cite{Cirelli:2023tnx} 
and CMB ($s$-wave) observations. 
For the $e^+e^-$ channel, the bound for 
$m_{\rm DM} > 5$ GeV is taken from \cite{Leane:2018kjk, Dutta:2022wdi} 
and it is a convolution of the limits from 
{\sc Ams} positron~\cite{PhysRevLett.113.121102, PhysRevLett.122.041102}, 
{\sc Fermi-Lat} dwarfs~\cite{Fermi-LAT:2016uux} 
and {\sc Hess} GC observations. 

\bigskip
 
Our main result, conveyed by {figs.~\ref{fig:sv_mx_limits_1} and \ref{fig:sv_mx_limits_2}}, 
is that, while the upper bounds (the red curves) on 
$\langle \sigma v \rangle$ obtained based on the existing 
$\gamma$-ray observations in MeV--GeV range 
remain mostly within the combined Indirect Detection exclusions 
(the green shaded regions) discussed above, 
the future MeV telescopes have the potential to probe large regions 
of the DM parameter space that are yet unexplored. 

Considering the projected sensitivities estimated with the 
Fisher matrix method, 
such regions for the $b\bar{b}$ and $W^+W^-$ annihilation channels 
extend up to a few hundreds of GeV 
(for {\sc Amego} and {\sc e-Astrogam}) or up to a 
few TeV (for {\sc Mast}), for thermal DM. 
For the leptonic channels the forecast sensitivities 
turn out to be even better. From fig.~\ref{fig:sv_mx_limits_1} 
we see that, for the $e^+e^-$ and $\mu^+\mu^-$ channels, the DM parameter space that 
remains unconstrained but lies within the reach of the future MeV telescopes 
can be extended up to the TeV or even the 
multi-TeV mass scale, for thermal DM. 
Such a statement is in general true for all the MeV telescopes considered here, 
although {\sc Mast} is expected to provide a comparatively better sensitivity. 
Considering a 100 GeV DM annihilating to $\mu^+\mu^-$, 
the Fisher forecasts for {\sc Amego} and {\sc e-Astrogam} 
reach a value of $\langle \sigma v \rangle$ which is 
almost a factor of 30 below the present experimental bound, 
while for {\sc Mast} this value can go down by an order of magnitude more.  

{Taking the rather simplified approach of 
the signal-to-noise ratio (SNR) method, the estimated sensitivities 
of the future telescopes extend even further. 
In {figs.~\ref{fig:sv_mx_limits_1} and \ref{fig:sv_mx_limits_2}}
this can be seen more prominently for $m_{\rm DM}$ up to the TeV scale 
for the $\mu^+\mu^-$ and $e^+e^-$ channels, 
and over almost all the DM mass range for the $b\bar{b}$ and $W^+W^-$ channels.} 

{Note that, compared to the simplified 
SNR approach, the Fisher method in general 
provides a more realistic estimate for  
signal normalization by accounting for statistical uncertainties in the backgrounds as well as 
the differences in the spectral shapes. In addition, it also takes 
into account correlations between background and signal. 
On the other hand, the SNR approach relies mainly on the 
comparison between the count of the number of photons from the signal to that from the total 
background (see Eq.~\ref{eq:SNR_method}). Hence, for a fiducial background model, 
the SNR is proportional to the signal number count which is 
the integration of the signal spectrum over the entire 
energy range of interest (e.g., $\sim0.2$ MeV -- 5 GeV for 
the MeV telescopes like {\sc Amego}). 
Now, in the cases where the total photon signal 
fills the entire photon energy window of interest (from low to high energies), 
such an integration of the signal spectrum over the energy leads to 
a large signal count and thus a stronger projection on the signal normalization 
compared to that obtained with the Fisher approach. For example, 
note from fig.~\ref{fig:DM_fluxes} that such a situation 
for the leptonic annihilation channels like $\mu^+\mu^-$ 
is more prominent for $m_{\rm DM}$ in the range 1 -- 100 GeV, 
thanks to the inclusion of low energy secondary 
photon spectra (mainly the ICS signals). 
As a result, for these mass scales, the SNR method 
yields better projections compared 
to the Fisher method, as can be seen by comparing the left and 
right panels of fig.~\ref{fig:sv_mx_limits_1}. 
As one moves towards the TeV mass scale, the corresponding contribution of 
the total signal becomes less efficient in filling the full energy window 
due to its fall at lower energies (even after including the secondaries) 
(see for the case $m_{\rm DM} = 1$ TeV in fig.~\ref{fig:DM_fluxes}). 
Hence, the total photon count also 
gets smaller, leading to a weaker SNR projection 
(which can be even weaker than that obtained with the Fisher method), 
as can be seen again by comparing the left and right panels of fig.~\ref{fig:sv_mx_limits_1}. 
On the other hand, for the hadronic channels, 
since the total signal (primary + secondaries) is softer and hence
covers the whole energy range interest in a somewhat uniform way 
for almost all DM masses (see, for example, the two plots for 
the $W^+W^-$ channel in fig.~\ref{fig:ISRF_variation_2}), 
the SNR method provides a better projection for all DM masses 
(as can be seen by comparing the left and right panels of fig.~\ref{fig:sv_mx_limits_2}).}

\medskip

Among the different MeV telescopes considered here, 
the forecast sensitivity of {\sc Mast} 
turns out to be comparatively better, mainly 
because of the large effective area 
estimated for this telescope (see fig.~\ref{fig:A_eff}). 
As can be seen from {figs.~\ref{fig:sv_mx_limits_1} and \ref{fig:sv_mx_limits_2}}, 
using such an instrument it could be possible to probe 
in the future a yet unconstrained 
thermally-annihilating DM model even at the 
scale of $\mathcal{O}(10)$ TeV.

\medskip 

MeV telescopes provide a good sensitivity for {\em leptonic} 
annihilation channels 
(like $\mu^+\mu^-$ and $e^+e^-$) mainly because these channels give rise, 
through cascades, to copious pairs of $e^\pm$, which in the case 
of usual GeV--TeV scale DM produce strong secondary signals 
as MeV $\gamma$-rays. 
This is illustrated in fig.~\ref{fig:sv_mx_limits_diff_components} (left panel), where 
the Fisher forecast projections for {\sc Amego} obtained considering 
the total DM signal  
are compared to those 
obtained considering only the prompt $\gamma$-ray signals. 
The prospects for the leptonic channels improve by up to 
two orders of magnitude when the secondary emission is considered, 
which shows clearly the importance of including such signals for weak-scale DM. 
For {\em hadronic} channels (right panel of fig.~\ref{fig:sv_mx_limits_diff_components}) 
the improvement is present but much reduced. 

\bigskip

Apart from the MeV $\gamma$-ray telescopes, there are various planned and 
proposed ground-based high-energy $\gamma$-ray telescopes such as 
{\sc Cta}~\cite{2013APh....43..171B} and {\sc Swgo}~\cite{Albert:2019afb},  
which are also expected to start operating in the forthcoming years. 
These telescopes are sensitive to $\gamma$-rays above $\sim50$ GeV. 
For example, the operating energy range for {\sc Cta} is 50 GeV -- 50 TeV, 
while for {\sc Swgo} it is 100 GeV -- 1 PeV. As a result, they 
can also play an important role in probing the annihilations 
of relatively heavy DM. In {figs.~\ref{fig:sv_mx_limits_1} and \ref{fig:sv_mx_limits_2}} 
we show the projected sensitivities of {\sc Cta} and {\sc Swgo} and compare them with 
those obtained for the MeV telescopes in this work. 
The solid and dashed black curves show the projections of {\sc Cta} 
for the observations towards the GC~\cite{Lefranc:2015pza} and the 
dwarf spheroidal (dSph) galaxy Ursa Major II~\cite{Lefranc:2016dgx}, 
respectively, for 500 hrs of observation time 
(that corresponds to $\sim5$ yrs of runtime \cite{PhysRevD.110.043003}). 
The dark blue solid curves show the projections of {\sc Swgo} for the observation towards the GC 
for 10 yrs~\cite{Viana:2019ucn}. The projections 
are shown here only for those channels that are available in the corresponding literature. 
As expected, among different targets, the highest sensitivities 
of these telescopes are achieved for the observation towards the GC. 
The high sensitivity of these telescopes for the energetic $\gamma$-rays or 
correspondingly for heavy DM is associated to their large effective areas based on the ground. 
As can be seen from {figs.~\ref{fig:sv_mx_limits_1} and \ref{fig:sv_mx_limits_2}}, 
while these high energy $\gamma$-ray telescopes may have a 
comparatively better sensitivity for 
a DM mass above a few 100 GeV (when compared to {\sc Amego} or {\sc e-Astrogam}) or a few TeV 
(when compared to {\sc Mast}), below such mass ranges 
the latter class of instruments provide the strongest sensitivities so far 
in probing DM signals. 
Therefore, one can say that 
the future space-based MeV $\gamma$-ray telescopes will very efficiently complement 
the ground-based high energy $\gamma$-ray instruments in the 
indirect searches for weak-scale DM. 

\medskip

{Note that in our work the projected sensitivities 
to DM annihilation from 
different future MeV telescopes are obtained using the two methods 
(SNR and Fisher) described in 
sections~\ref{sec:SNR_method} and \ref{sec:Fisher_method}. 
These methods are efficient in obtaining the projections for DM in the sense that 
they focus on comparing the putative 
DM signal to the astrophysical backgrounds by 
taking into account the statistical fluctuations in the background model as well as 
the correlations between backgrounds and signal (applicable for the Fisher analysis). 
However, they do not account directly for the systematic uncertainties 
in the background model itself, which can be dominant over the statistical fluctuations 
and are expected to be at the level of 15\% for the 
considered backgrounds \cite{Boddy:2015efa, Strong:2004de}. 
Since our analyses (based on SNR and Fisher forecasts) 
use the square-root of the background photon flux, 
the addition of a systematic uncertainty to our fiducial background at a 
level of 15\% is expected to weaken our projections 
by $\sim7$\% only (see also \cite{ODonnell:2024aaw} for a similar discussion). 
In order to verify this for both the methods, we re-estimate some of the projections 
shown in figs.~\ref{fig:sv_mx_limits_1} and \ref{fig:sv_mx_limits_2} 
by conservatively enhancing each of the fiducial background components 
(discussed in section \ref{sec:background}) by a factor of 2 (at all photon energies), 
and verified that indeed the projections are weakened {\it maximally} 
by $\sim40$\% compared to the fiducial ones. 
For a somewhat similar quantitative discussion about this, see 
for example \cite{Negro:2021urm}.}

\subsection{{Discussion on the effects of 
propagation of $e^\pm$ in the Galaxy}}
\label{sec:propagation}

\begin{figure*}[ht!]
\hspace{-1mm}\includegraphics[width=0.5\textwidth]
{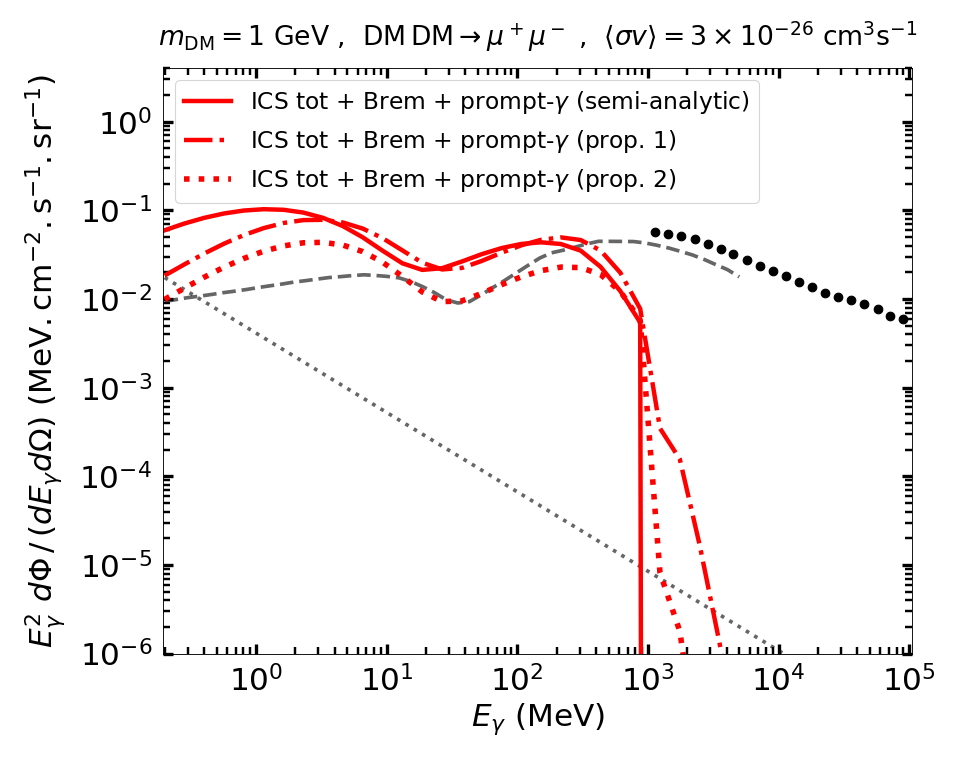}
\includegraphics[width=0.5\textwidth]
{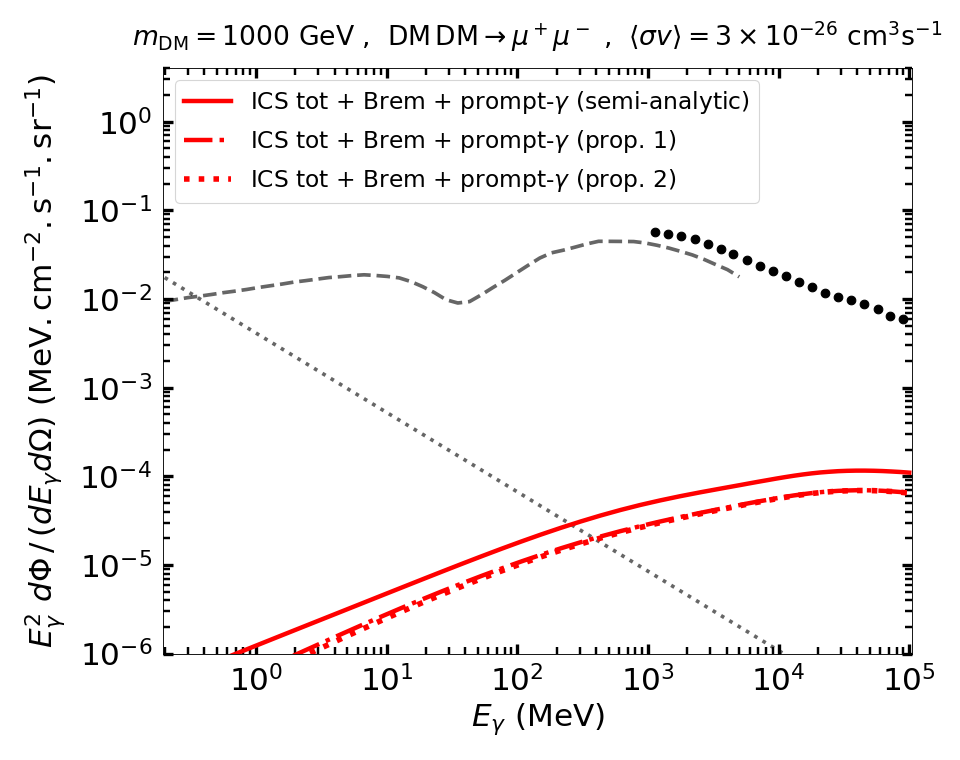}
\vspace{-6mm}
\caption{{\em {\bfseries Effects of the propagation} of $e^\pm$. 
The red solid curves correspond to the 
total photon flux (ICS + brem + prompt-$\gamma$) shown in fig.~\ref{fig:DM_fluxes}, where the secondaries 
are obtained with the $e^\pm$ distribution resulting from the semi-analytic approach (Eq.~\ref{eq:dnedE}). 
The red broken curves show the same total photon flux, now computed with the $e^\pm$ distribution 
in the Galaxy resulting from solving the full propagation equation with \texttt{DRAGON2}~\cite{Evoli:2016xgn, Evoli:2017vim}. 
In particular, the red dashed-dotted and dotted curves correspond to the two different propagation models 
discussed in the text. The backgrounds (gray lines) and 
the {\sc Fermi} data (black points) are shown for illustration and are the same as the ones presented in fig.~\ref{fig:DM_fluxes}.}}
\label{fig:DM_fluxes_fullTransport}
\end{figure*}

{In this sub-section we discuss the effects of considering the full 
propagation of DM-induced $e^\pm$ in the Galaxy. 
In fig.~\ref{fig:DM_fluxes_fullTransport} we show these effects on the 
DM induced photon signal for two values of the DM mass, 
$m_{\rm DM} = 1$ GeV and 1 TeV, considering the $\mu^+\mu^-$ annihilation channel. 
The red solid curves correspond to the sum of DM-induced 
ICS, bremsstrahlung and prompt $\gamma$-ray fluxes shown 
in fig.~\ref{fig:DM_fluxes}, where the $e^\pm$ distribution required 
to estimate the secondary signals in the Galaxy 
were obtained using eq.~\eqref{eq:dnedE}. 
The red broken curves, on the other hand, correspond to the sum of 
similar quantities, but the $e^\pm$ distribution 
here are obtained by solving the full Galactic 
propagation that includes different processes, such as 
spatial diffusion, advection, convection, 
re-acceleration, radiative energy losses and various nuclear processes; 
see for example eq.~2.1 of \cite{Evoli:2016xgn} or \cite{Evoli:2017vim}. 
In order to solve this we use the package 
\texttt{DRAGON2}~\cite{Evoli:2016xgn, Evoli:2017vim}~\footnote{{\texttt{DRAGON2} (used here in the context of estimating the galactic DM signals) 
is similar in scope to \texttt{GALPROP} (used in the literature to estimate different Galactic backgrounds mentioned above). Both are designed to cover most of the relevant processes involving galactic 
cosmic-rays and their secondary products, over a wide energy range. 
\texttt{DRAGON2} and \texttt{GALPROP} implement similar algorithms 
to solve the galactic propagation 
equation of charged par-} {ticles 
such as $e^\pm$'s, see \cite{Evoli:2016xgn} and \cite{Strong:1998pw, galprop_v54}. 
Moreover, the energy-loss processes, ISRF models, 
gas maps and various nuclear processes implemented in \texttt{DRAGON2} 
are quite similar to the ones used in \texttt{GALPROP}. 
See \cite{Evoli:2016xgn} for a detailed discussion about these.}}. 
The magnetic field is assumed to be the default model 
implemented in \texttt{DRAGON2}.  
The DM profile is considered to be the same one used in fig.~\ref{fig:DM_fluxes}.} 

{The red dashed-dotted and dotted curves 
in fig.~\ref{fig:DM_fluxes_fullTransport} 
correspond to two Galactic propagation models, 
named here as `prop.~1' and `prop.~2', respectively. 
These models have been used in the past for the study related to 
the secondary photon emissions towards the GC.}

{\underline{\bf Model `prop.~1':} it refers to the ``model z04LMS" from \cite{2010ApJ...722L..58S} 
(see their Table 1) which was adopted in \cite{2011crpa.conf..473S} 
to estimate the different Galactic photon backgrounds 
(i.e., bremsstrahlung, $\pi^0$ and ICS components) 
mentioned in section~\ref{sec:background} and 
considered in the present work. 
In such model the spatial diffusion is parameterized as a function 
of the propagating particle rigidity ($\rho$) as 
$D (\rho) = D_0 \, \beta^\eta \, (\rho / 4 \, {\rm GV})^\delta$, 
with $\beta$ as the dimensionless particle velocity and $\eta = 1$. 
The values of $D_0$ and $\delta$ are $5.8\times10^{28}$ ${\rm cm^2 s^{-1}}$ and 
0.33, respectively. The Alfvén velocity 
(related to the re-acceleration of the particles) is $v_A = 30$ km/s. 
The size of the diffusion zone (beyond which the $e^\pm$ escape freely) 
is the same one mentioned in section~\ref{sec:DMsignal} and used in our other results 
(i.e., $R_{\rm Gal} = 20 \, {\rm kpc}$ and $L_{\rm Gal} = 4 \, {\rm kpc}$).}

\begin{figure*}[ht!]
\includegraphics[width=0.5\textwidth]{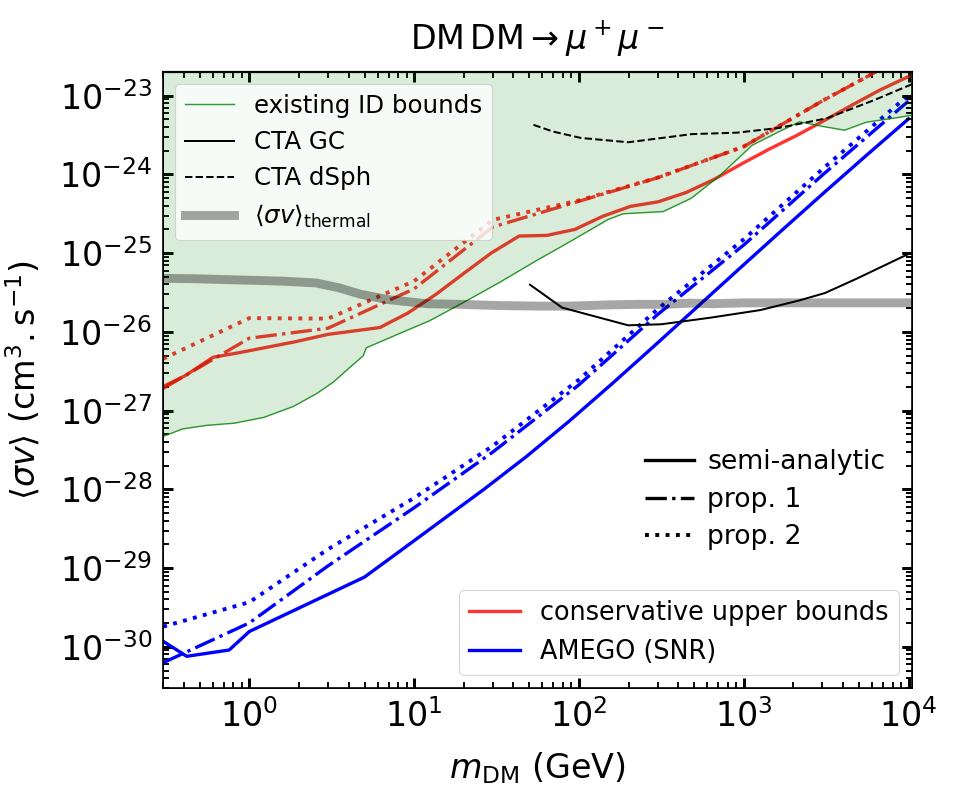}
\includegraphics[width=0.5\textwidth]{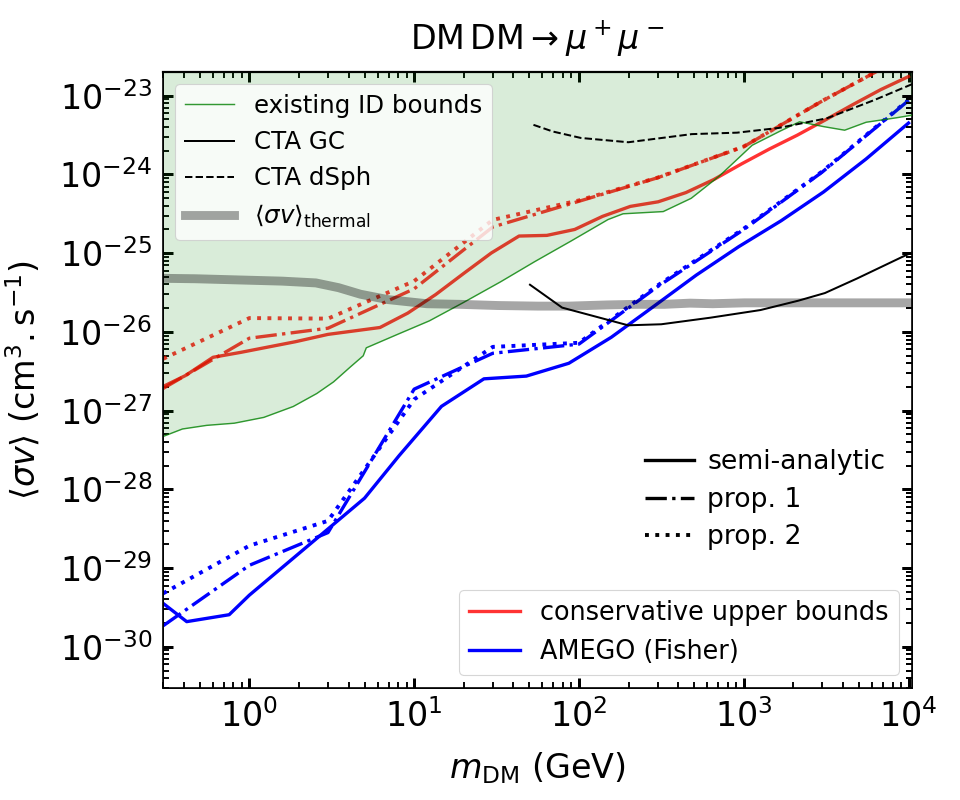}
\caption{{\em {\bfseries Effects of the propagation of 
$e^\pm$ on the estimated bounds and projections} 
are shown for the $\mu^+\mu^-$ annihilation channel. 
The projections are shown for {\sc Amego} and are obtained using 
the two statistical approaches discussed in sections~\ref{sec:SNR_method} (left panel) 
and \ref{sec:Fisher_method} (right panel). The red and blue solid curves are respectively 
the same bounds and projections shown in fig.~\ref{fig:sv_mx_limits_1} 
and correspond to the case where 
the $e^\pm$ distribution in the Galaxy is obtained using the semi-analytic method 
described by Eq.~\eqref{eq:dnedE}. The red and blue broken curves, on the other hand, 
show the results obtained with the 
full propagation of $e^\pm$ in the Galaxy, for the two models discussed in the text.}}
\label{fig:DM_limits_fullTransport}
\end{figure*}  

{\underline{\bf Model `prop.~2':} it refers to the ``model A" adopted in \cite{Calore:2014xka} 
(see their Table 2) to estimate the flux for the above-mentioned Galactic backgrounds. 
This model was used in \cite{ODonnell:2024aaw} to rescale the 
Galactic backgrounds to the region $10^{\circ}$ cone around 
the GC which is our target ROI. 
For this model, $R_{\rm Gal} = 20 \, {\rm kpc}$, 
$L_{\rm Gal} = 4 \, {\rm kpc}$, $D_0 = 5.0\times10^{28}$ ${\rm cm^2 s^{-1}}$, 
$\eta=0$, $\delta = 0.33$, $v_A = 32.7$ km/s. 
In addition, this model includes a somewhat realistic convective wind velocity 
in terms of its gradient perpendicular to the Galactic plane, $dv/dz = 50$ km/s/kpc. 
Notice that Model `prop.~2' is actually very similar to `prop.~1' for relativistic $e^\pm$; the only difference is that `prop.~2' includes a convective wind which the `prop.~1' does not.} 

{Fig. \ref{fig:DM_fluxes_fullTransport} shows that, 
considering the full effects of propagation of $e^\pm$ in the Galaxy,  
the total DM signal across the GeV--TeV DM mass range 
can be suppressed {\it at most} by a factor of few 
(depending on the propagation models) in the entire photon energy range of interest.} 

{To show this more prominently for the full DM mass range considered, 
we show in Fig. \ref{fig:DM_limits_fullTransport} the variations in the 
upper-bounds and the projections of the MeV telescopes in the 
$\langle \sigma v \rangle - m_{\rm DM}$ plane due to the effects of the 
Galactic propagation of $e^\pm$. 
For definiteness, this figure is shown only for 
the $\mu^+\mu^-$ annihilation channel and for the MeV telescope {\sc Amego}. The left and right panels correspond to the two statistical approaches 
discussed in sections~\ref{sec:SNR_method} and \ref{sec:Fisher_method}, respectively. 
The solid red and blue curves refer to respectively 
the same bounds and projections shown in fig.~\ref{fig:sv_mx_limits_1} 
and correspond to the case where the $e^\pm$ distribution in the Galaxy 
is obtained using the semi-analytic method described by Eq.~\ref{eq:dnedE}. 
On the other hand, the dashed-dotted (dotted) red and blue curves show the results 
estimated involving the full propagation of $e^\pm$ in the Galaxy with the 
propagation model `prop.~1' (`prop.~2') discussed above.} 

{From Fig. \ref{fig:DM_limits_fullTransport} one can see that 
including a detailed and involved numerical simulation 
of the propagation of $e^\pm$ in the Galaxy under 
some commonly used propagation models 
(that are used to estimate the Galactic photon backgrounds towards the GC) 
the bounds and the projections of MeV telescopes on the weak-scale 
DM parameter space can be weakened {\it maximally} by a factor $\sim2-3$ 
depending on the DM mass. 
Hence, the discussion of this section indeed shows that 
the conclusions based on the results obtained adopting 
the semi-analytic approach to estimate the Galactic distribution of DM 
induced $e^\pm$ remain the same.} 

{Using propagation models different 
than the `prop.~1' or `prop.~2' used here is expected to change not only the DM induced signals 
but also to modify the Galactic backgrounds 
(composed of secondary radiations) 
in a similar fashion, leaving the constraints on DM annihilation 
in the same ballpark. For example, assuming a propagation scenario 
that enhances the signal flux, say, by a factor of a few 
(like in the case of the semi-analytic approach), 
one may expect the background flux to be also modified 
in a similar way over the energy range of interest. 
Since the analysis for obtaining the projections on DM 
(e.g., the signal-to-noise ratio in Eq.~\eqref{eq:SNR_method}) 
relies on the comparison of the signal to the fluctuation in the background flux 
(appearing as the square-root), the effects of both enhancements are expected to 
mitigate each other (more precisely, actually, the signal-to-noise 
ratio should be enhanced slightly, making the projections on the 
DM annihilation slightly stronger). 
There is also a possibility of translating such variations into the systematics. 
Hence, in order to be less dependent on different propagation parameters 
and to keep full control of our computations, we adopted the semi-analytic approach 
described in sec.~\ref{sec:DMsignal} to obtain our main results (presented 
in figs.~\ref{fig:sv_mx_limits_1} and \ref{fig:sv_mx_limits_2}), 
and leave a more detailed numerical study involving 
different propagation models self-consistently applied to signals 
as well as to backgrounds for future work.}

\section{Conclusions}
\label{sec:conclusions}

In this work, we have explored the sensitivity of future MeV telescopes to the photon signals produced by the annihilation of weak-scale DM particles, focusing in particular on the galactic center region. 
The central idea of our study is that these telescopes will be sensitive to the low-energy secondary emissions (essentially ICS and bremsstrahlung $\gamma$-rays produced by the DM-induced electrons and positrons) and will therefore be able to complement the high-energy $\gamma$-ray experiments, which are instead insensitive to the DM prompt emission. We focused on three representative planned experiments ({\sc Amego}, {\sc e-Astrogam} and {\sc Mast}) and on a few representative DM annihilation channels. We adopted for definiteness a NFW DM galactic distribution, a streamlined treatment of the galactic propagation of $e^\pm$ and standard assumptions for the astrophysical environment 
(but we investigate in detail the impact of varying all these ingredients 
{in section~\ref{sec:propagation} and} 
in Appendix \ref{sec:astrophysical_models}).
We used two forecast approaches: a simple signal-over-noise ratio criterion and the more refined Fisher matrix method. 

Our results are very promising. We find that, thanks to the fact that the secondary emissions significantly enhance the MeV signals of weak-scale DM annihilations, the MeV telescopes will explore a wide area of the $m_{\rm DM}-\langle \sigma v \rangle$ parameter space. {Figs.~\ref{fig:sv_mx_limits_1} and \ref{fig:sv_mx_limits_2}} summarize 
our main results. We find that an experiment like {\sc Mast}, for an observation time of about 3 years, 
will be able to probe thermally-annihilating DM up to a few TeV's of mass, 
the details depending on the annihilation channel and the analysis method. 
This is comparable to the reach of future high-energy telescopes such as {\sc Cta} and {\sc Swgo}. 
For lower masses, the future MeV telescopes could be sensitive to DM annihilation cross sections that are 
2 to 3 orders of magnitude smaller than the current bounds, again with the details 
depending on the channel and the analysis. 

Summarizing, it is remarkable that MeV telescopes will be able to probe TeV DM, possibly competing with TeV telescopes, because of the significant leverage of secondary emissions. This adds an important tool in the continuing quest for the indirect detection of weak-scale DM.

\small
\subsubsection*{Acknowledgments}
\footnotesize
The authors acknowledge useful discussions with HaoYu Xie. They acknowledge the hospitality of the Institut d'Astrophysique de Paris ({\sc Iap}) where part of this work was done. M.C.~also acknowledges the hospitality of the Flatiron Institute of the Simons Foundation and New York University. 

\noindent Funding and research infrastructure acknowledgments: 
{\sc Cnrs} grant {\sl DaCo:~Dark Connections}; research grant {\sl DaCoSMiG} from the {\sc 4eu+} Alliance (including Sorbonne Université); Institut Henri Poincaré (UAR 839 CNRS-Sorbonne Université) and LabEx CARMIN (ANR-10-LABX-59-01).

\normalsize
\appendix

\section{Impact of astrophysical uncertainties}
\label{sec:astrophysical_models}

\subsection{DM halo profile}
\label{sec:appendix_DM_halo}

\begin{figure*}[ht!]
\centering
\includegraphics[width=0.49\textwidth]{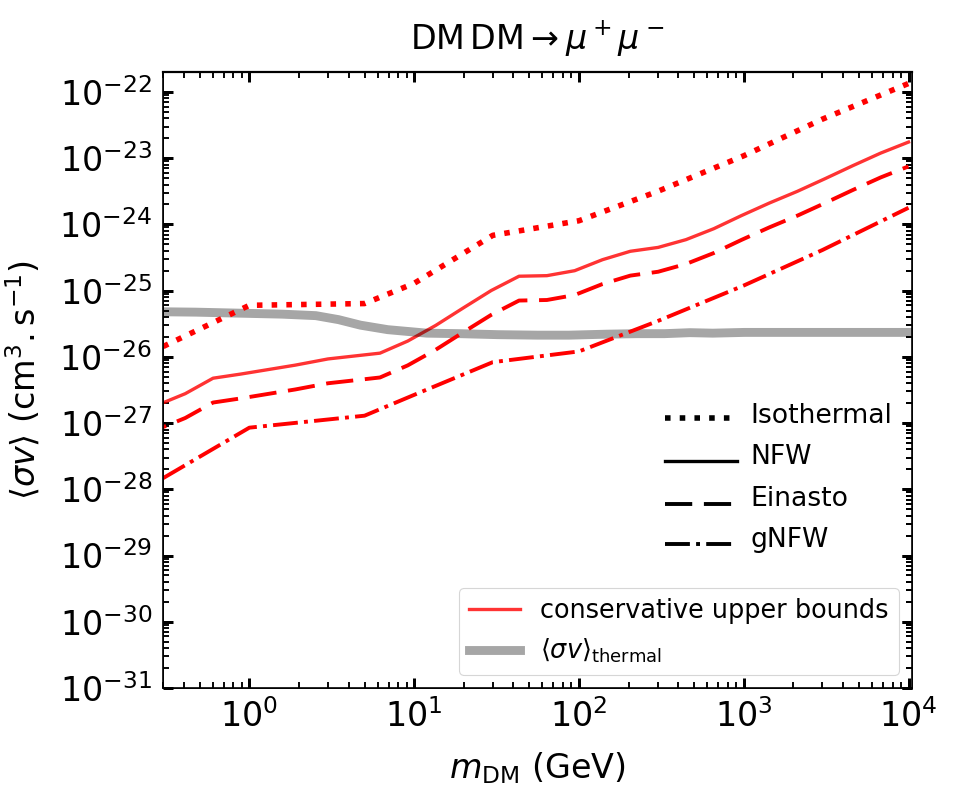} \\ 
\vspace{-1mm}
\includegraphics[width=0.49\textwidth]{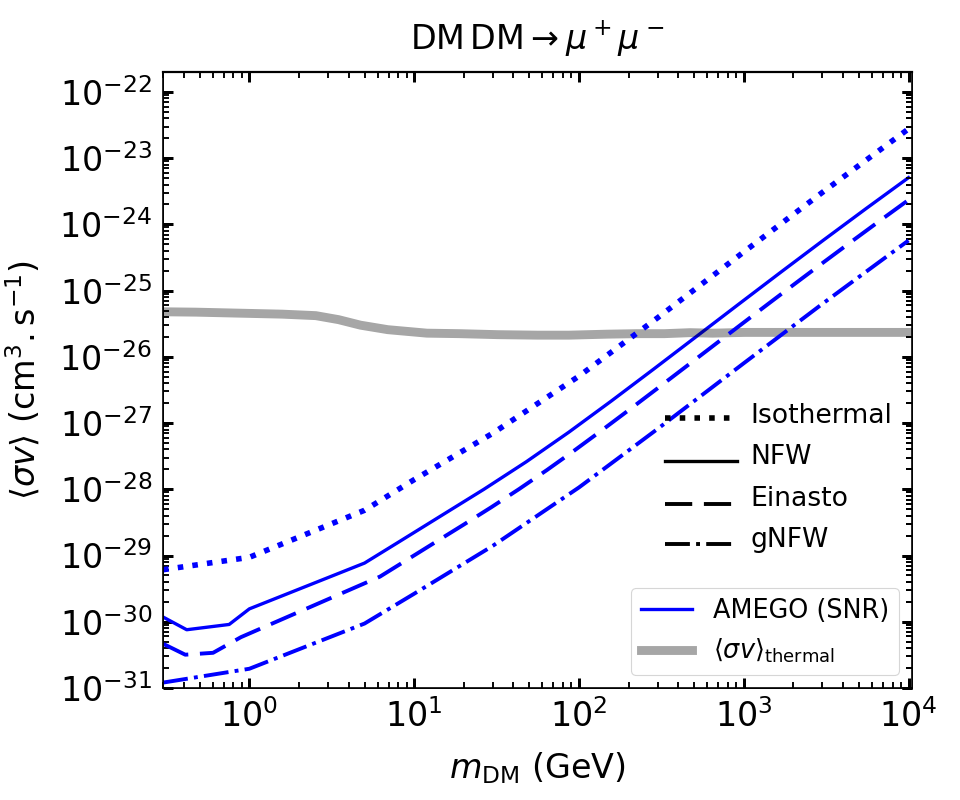}
\includegraphics[width=0.49\textwidth]{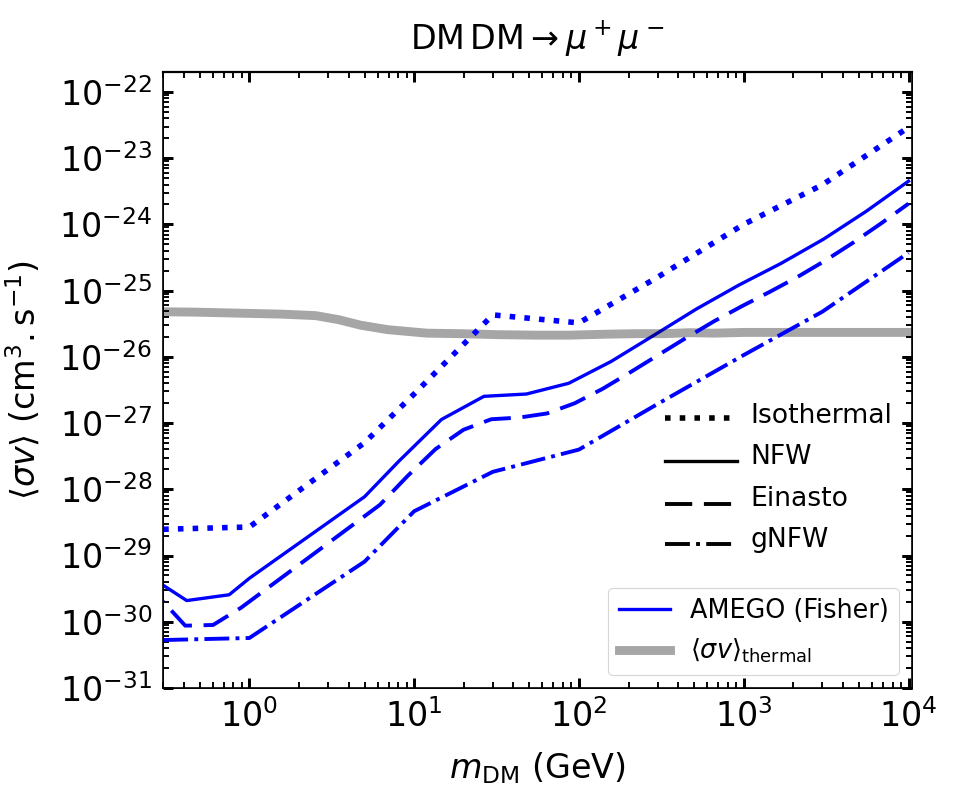}
\caption{{\em Comparison of the bounds (upper panel) and 
projections for {\sc Amego} (bottom panels) 
obtained with {\bfseries different galactic DM halo profiles}. 
The projections are shown for the two statistical approaches discussed in 
sections~\ref{sec:SNR_method} (bottom left) and \ref{sec:Fisher_method} (bottom right). 
The colored solid curves correspond to the NFW profile and are the same as  
the ones shown in fig.~\ref{fig:sv_mx_limits_1}. The dashed, dashed-dotted and dotted curves 
correspond to Einasto, generalized NFW (gNFW) and Isothermal profiles, respectively. See the text for details.}}
\label{fig:DM_profile_variation}
\end{figure*}

Fig.~\ref{fig:DM_profile_variation} shows how our results, {i.e. the bounds (top panel) and 
projections (bottom panels),} vary with 
different choices for the galactic halo DM profile. 
The colored solid curves correspond to the NFW profile (eq.~\eqref{eq:rho_profile})
with the parameters $\rho_0$ and $r_s$ corresponding to the 
fitted parameters (the central values) obtained in~\cite{2019JCAP...10..037D} 
for the baryonic model B2, as used in all other results of this work. 
The dashed and dashed-dotted curves show, respectively, the results 
obtained using the Einasto profile~\cite{1965TrAlm...5...87E} 
\begin{equation}
\rho^{\rm Ein}_{\rm DM}(r) = \rho_0 \, {\rm exp} \left\lbrace -\frac{2}{\alpha} 
\left(\left(\frac{r}{r_s}\right)^\alpha - 1\right) \right\rbrace \, ,
\label{eq:rho_profile_Einasto}
\end{equation}
and a generalized NFW (gNFW) profile
\begin{equation}
\rho^{\rm gNFW}_{\rm DM}(r) = \frac{\rho_0}{\left(\frac{r}{r_s}\right)^\gamma 
\left(1 + \frac{r}{r_s}\right)^{3-\gamma}} \, .
\label{eq:rho_profile_gNFW}
\end{equation}
The corresponding profile parameters are set to their 
central values obtained in the fit in~\cite{2019JCAP...10..037D} 
using the same baryonic model B2 used for the NFW profile. 
The values of $\alpha$ and $\gamma$ corresponding to 
the Einasto and gNFW profiles are $\alpha = 0.18$ and 
$\gamma = 1.3$, respectively. 
In fig.~\ref{fig:DM_profile_variation} we also show, with dotted curves, 
the results obtained using a truncated isothermal (cored) profile
\begin{equation}
\rho^{\rm Iso}_{\rm DM}(r) = 
\frac{\rho_0}{1 + \left(\frac{r}{r_s}\right)^2} \, ,
\label{eq:rho_profile_Iso}
\end{equation}
with the corresponding parameters tabulated in~\cite{Cirelli:2024ssz}.  

Form fig.~\ref{fig:DM_profile_variation} we see that, 
while considering Einasto and gNFW profiles the bounds 
and projections presented in fig.~\ref{fig:sv_mx_limits_1}  
can strengthen by a factor of 2 and by an order of magnitude, respectively, 
considering the Isothermal profile they loose by a factor of 10.

\subsection{ISRF model}
\label{sec:appendix_ISRF_model}

\begin{figure*}[ht!] 
\centering
\includegraphics[width=0.496\textwidth]{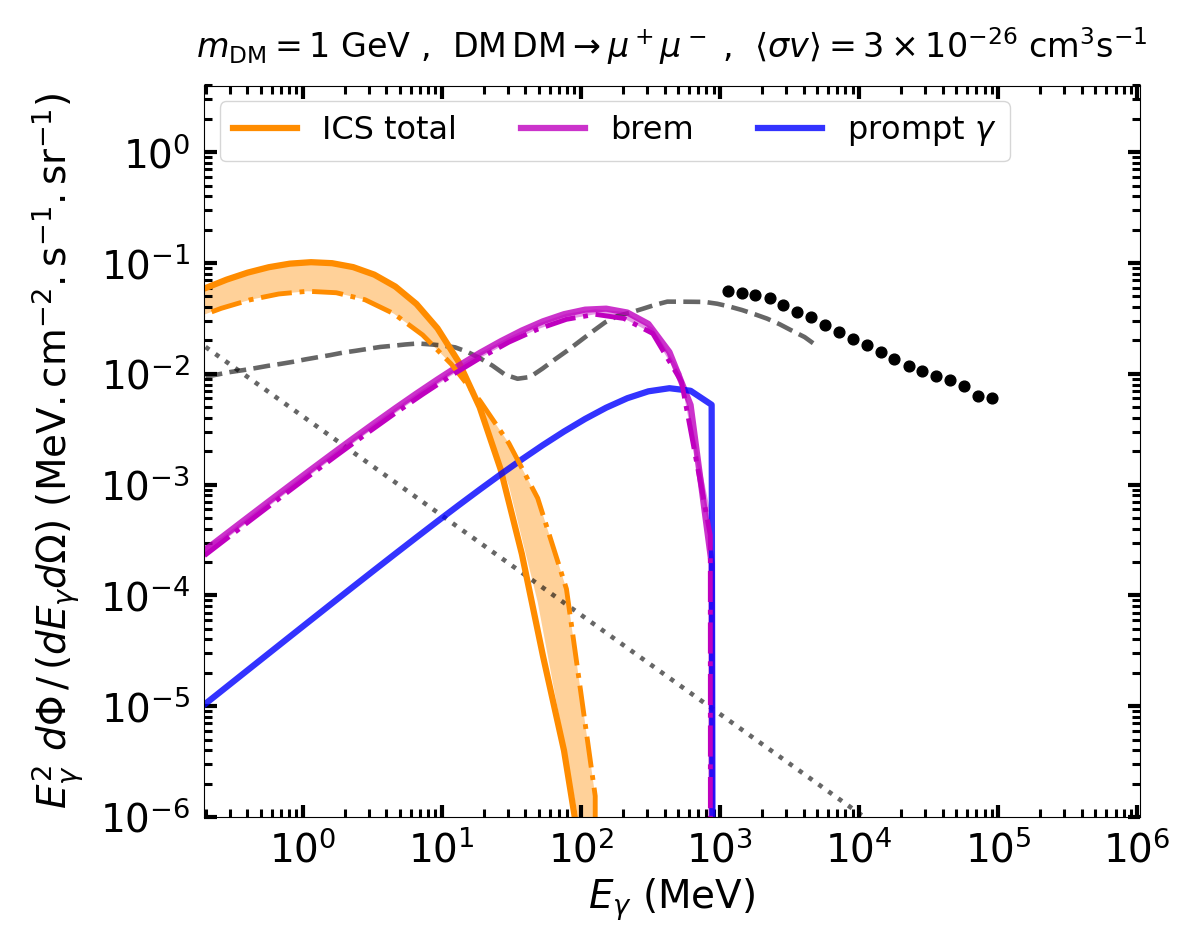}
\includegraphics[width=0.496\textwidth]{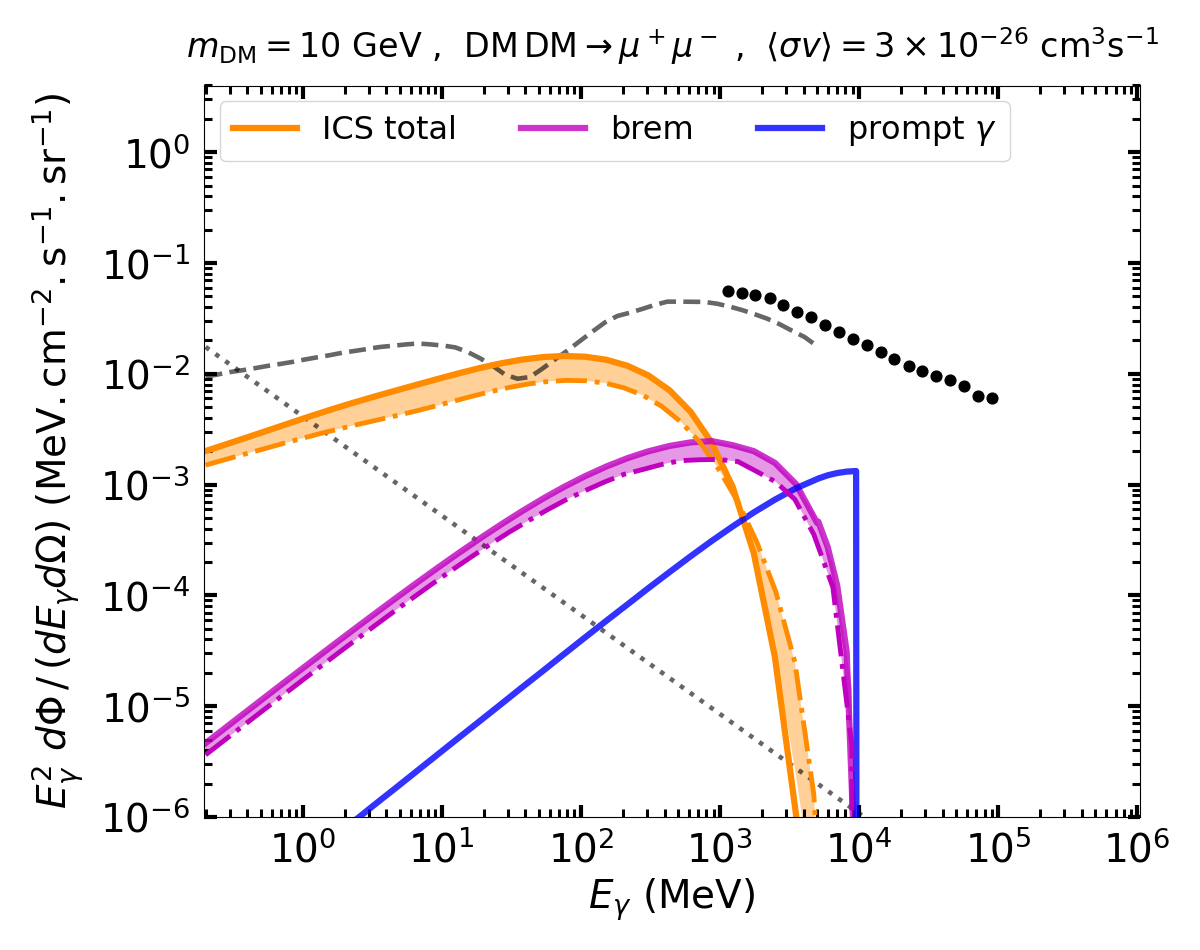}
\includegraphics[width=0.496\textwidth]{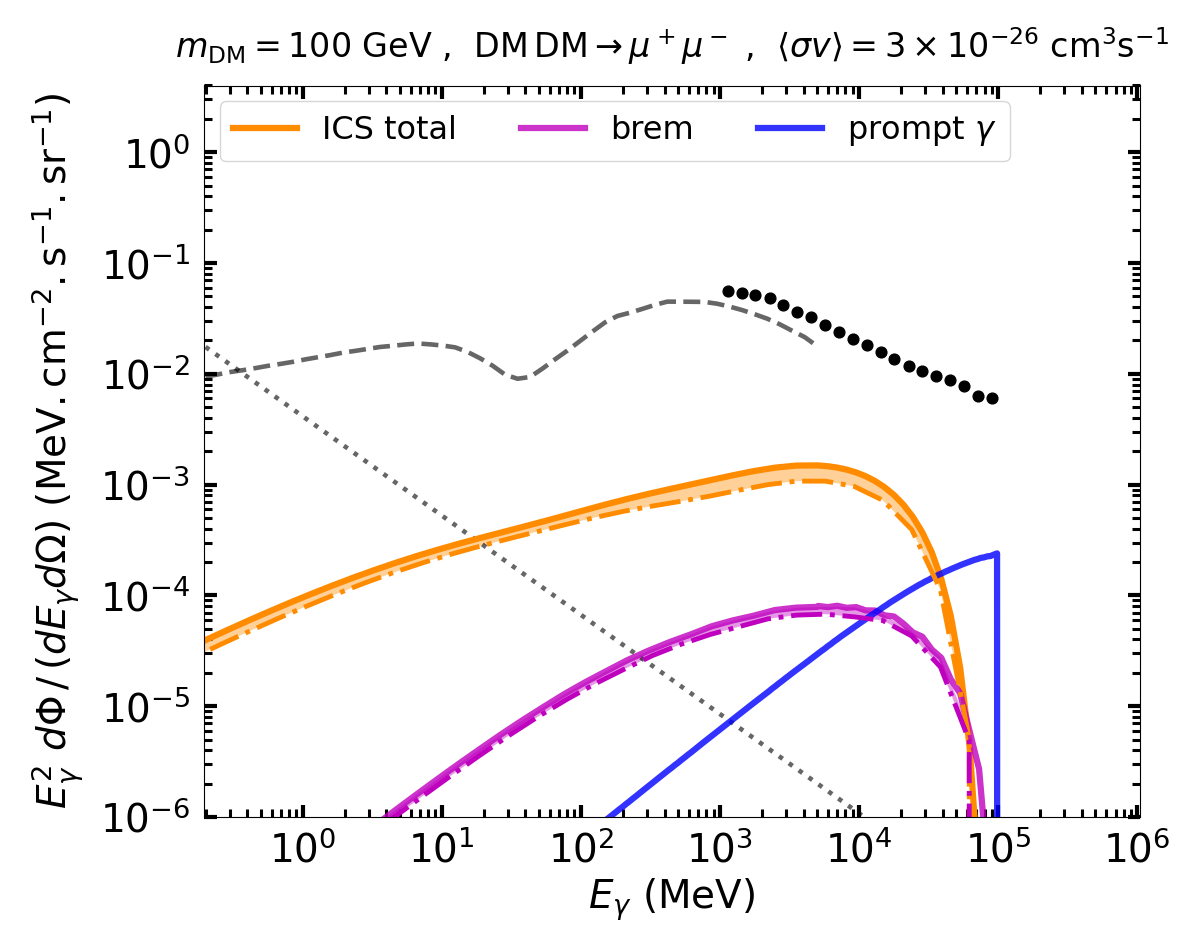}
\caption{\em DM-induced photon signals considering {\bfseries different ISRF models}. 
{The results are shown for $m_{\rm DM} =$ 1, 10 and 100 GeV considering 
the $\mu^+\mu^-$ annihilation channel.}
The dashed-dotted curves (for the ICS and bremsstrahlung signals) 
are obtained considering the ISRF model 
``Porter2006" (see \cite{Evoli:2016xgn} for a detailed discussion).  
The solid curves on the other hand 
correspond to the ISRF model discussed in 
sec.~\ref{sec:secondary_ICS} and are the same as the ones 
shown in fig.~\ref{fig:DM_fluxes}.
}
\label{fig:ISRF_variation}
\end{figure*}

\begin{figure*}[ht!] 
\includegraphics[width=0.5\textwidth]{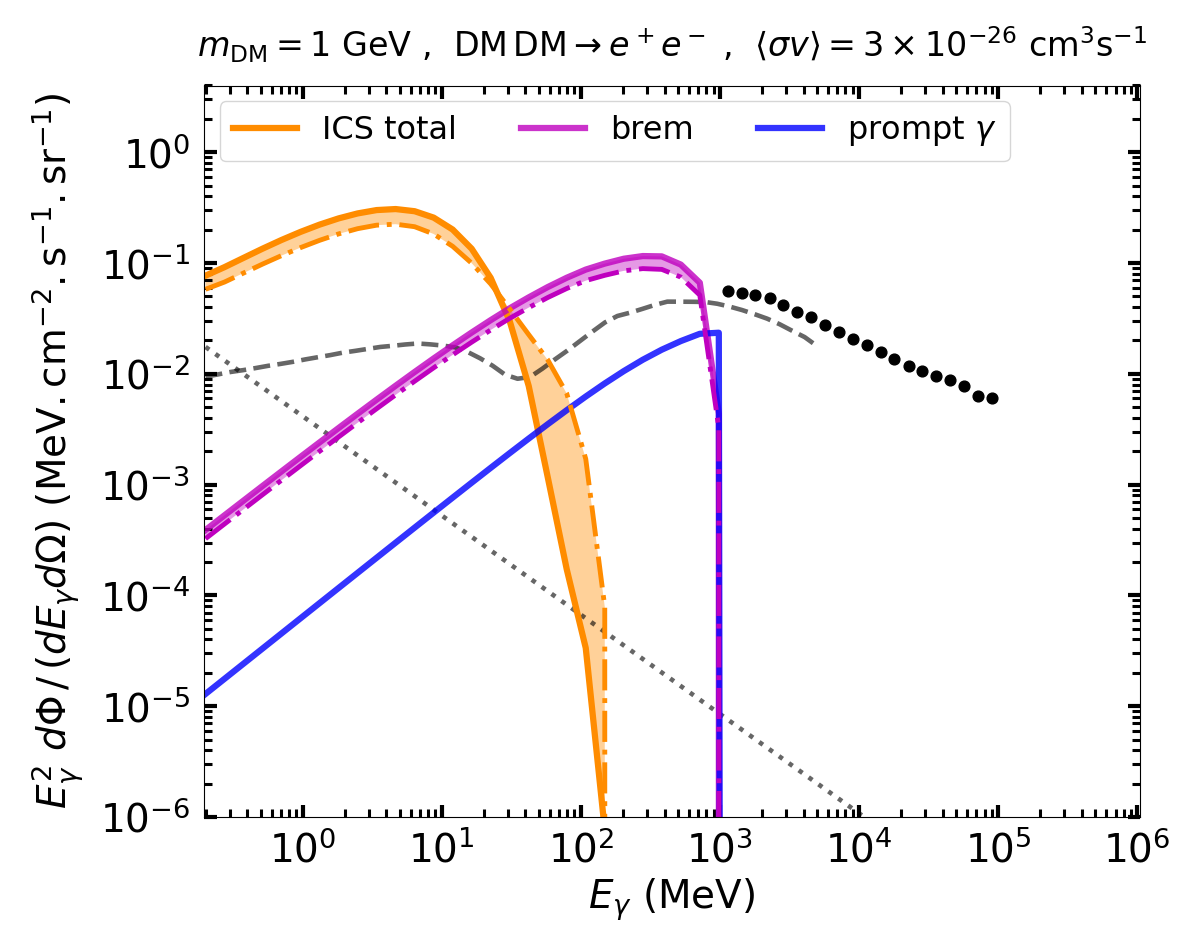}
\includegraphics[width=0.5\textwidth]{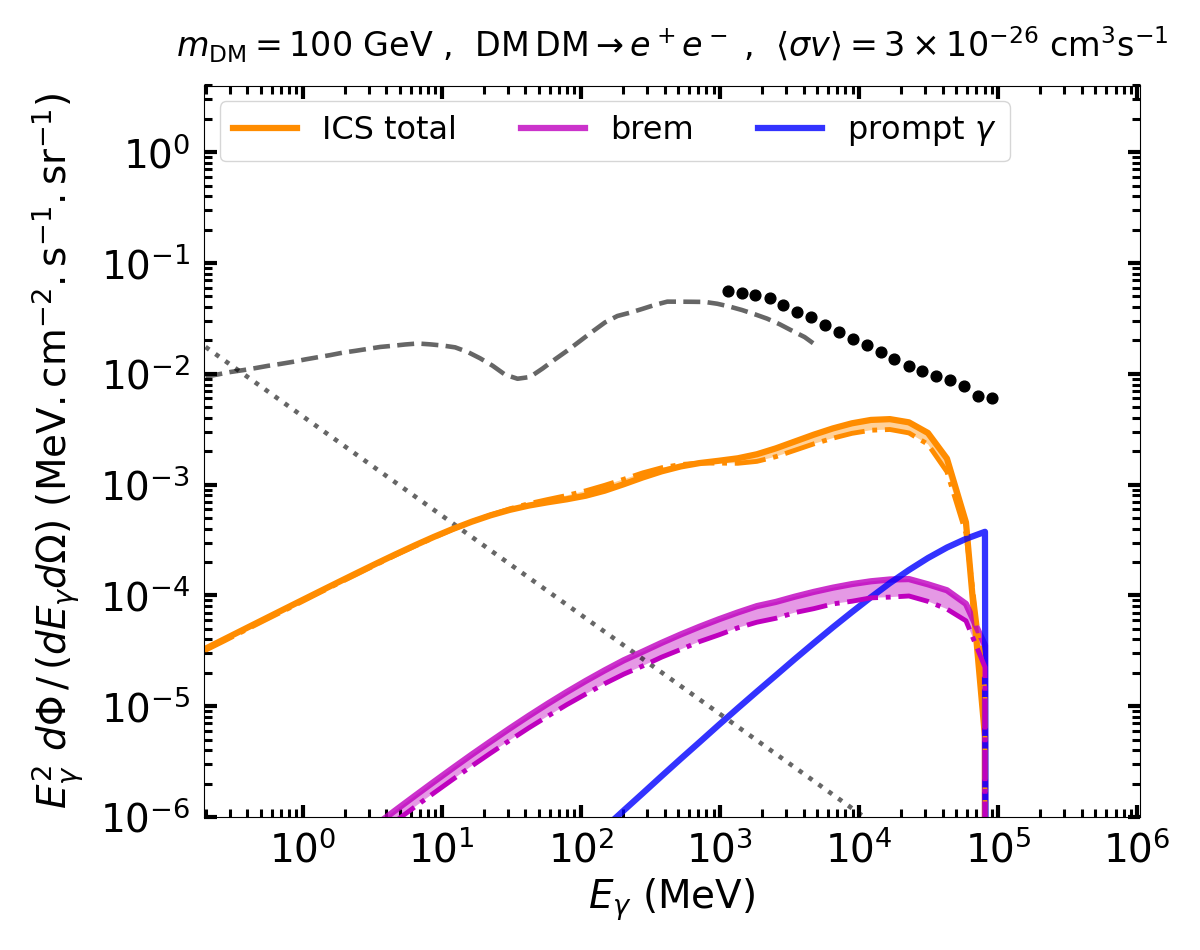}
\includegraphics[width=0.5\textwidth]{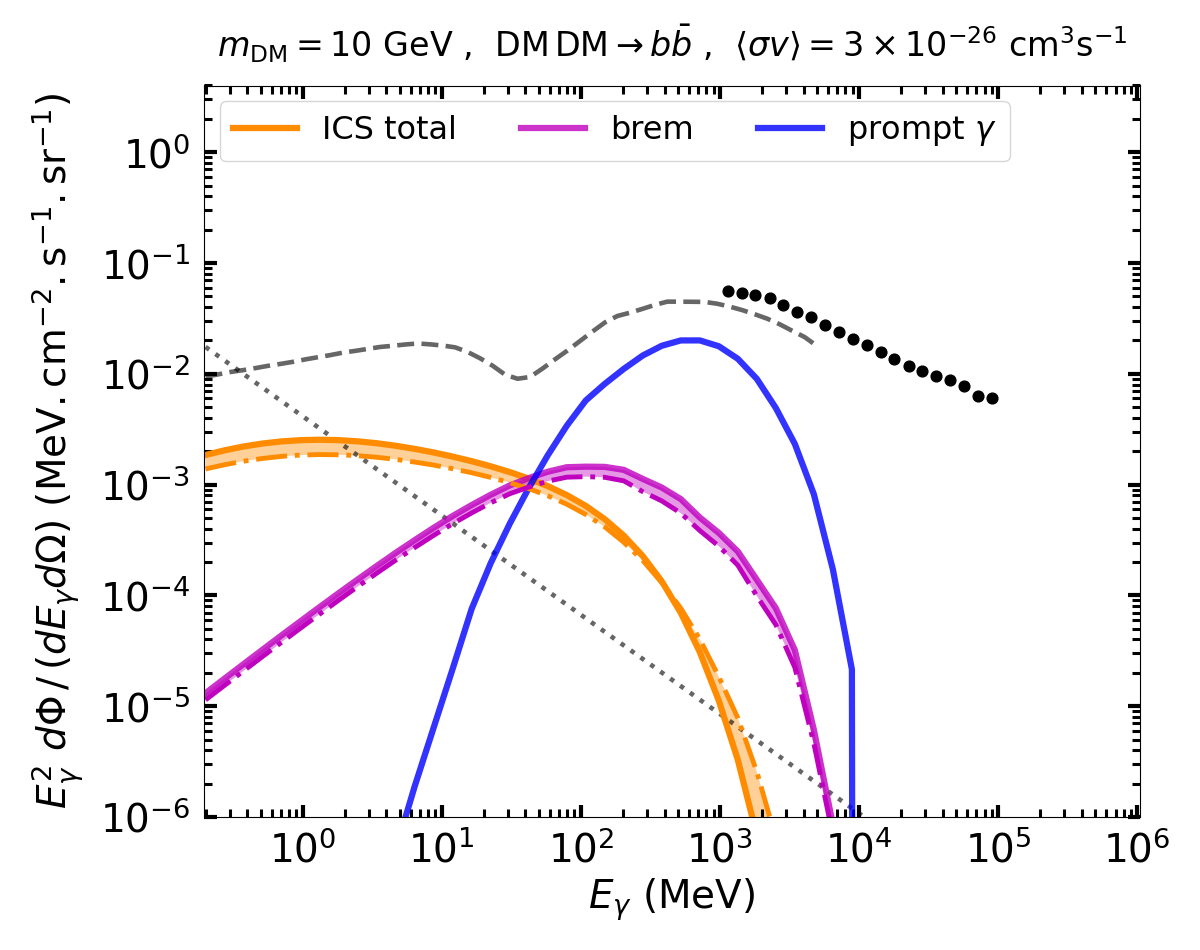}
\includegraphics[width=0.5\textwidth]{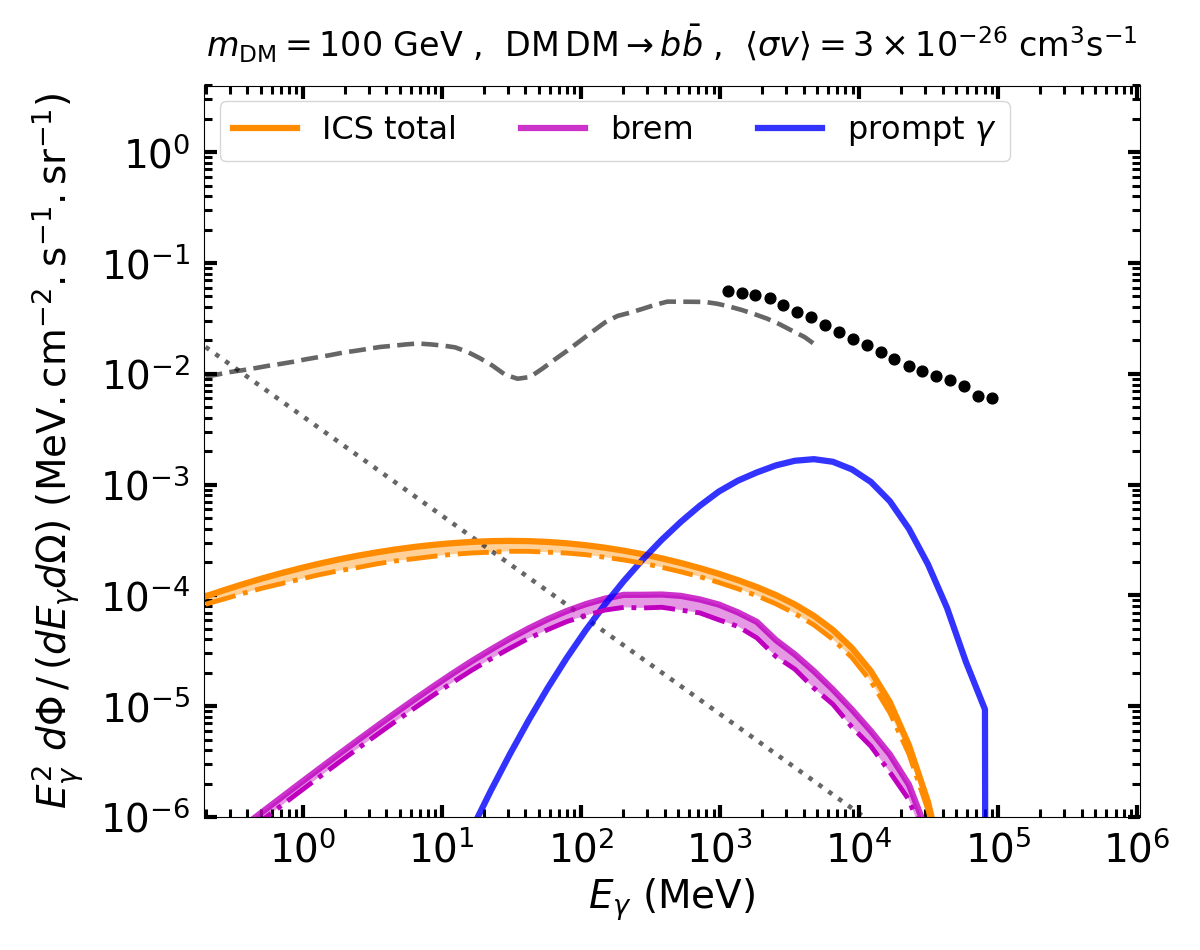}
\includegraphics[width=0.5\textwidth]{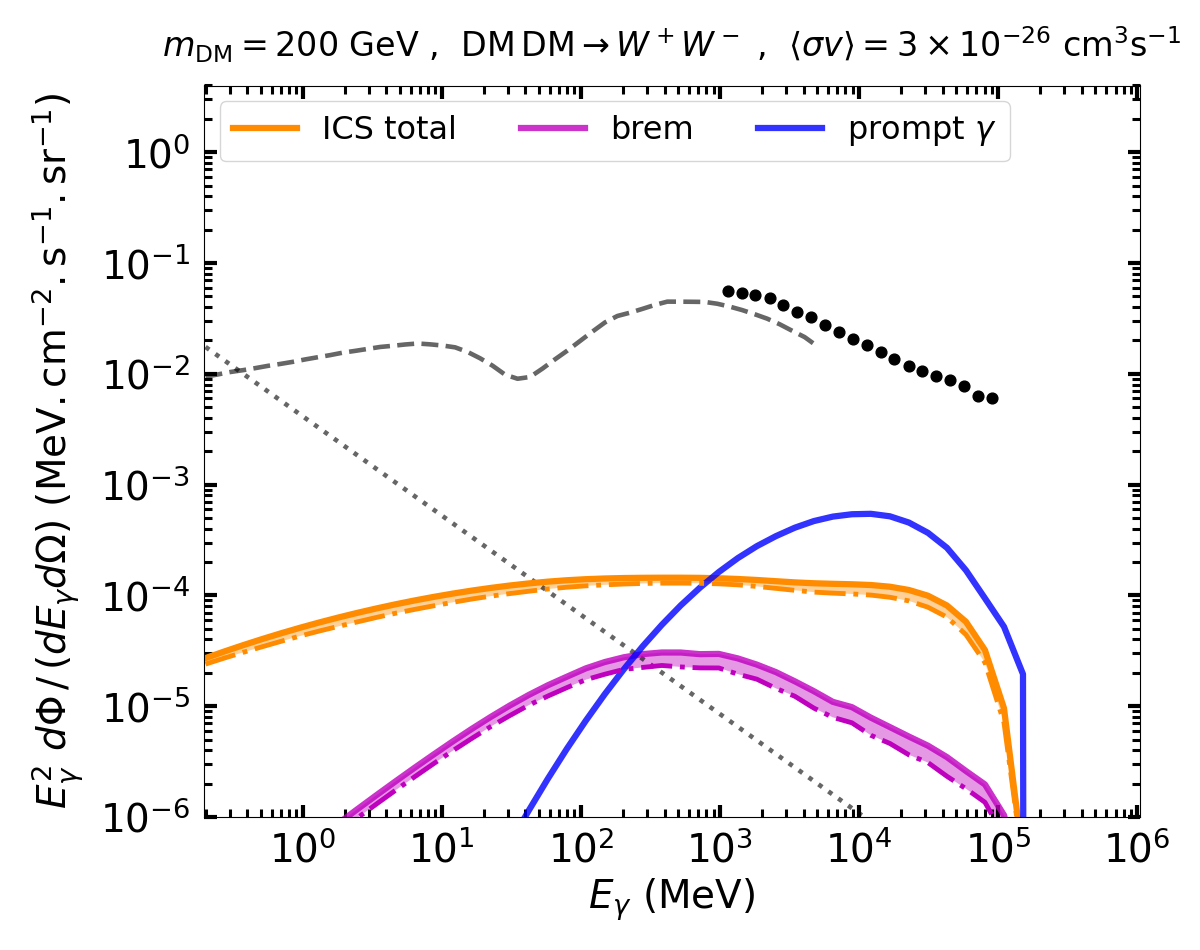}
\includegraphics[width=0.5\textwidth]{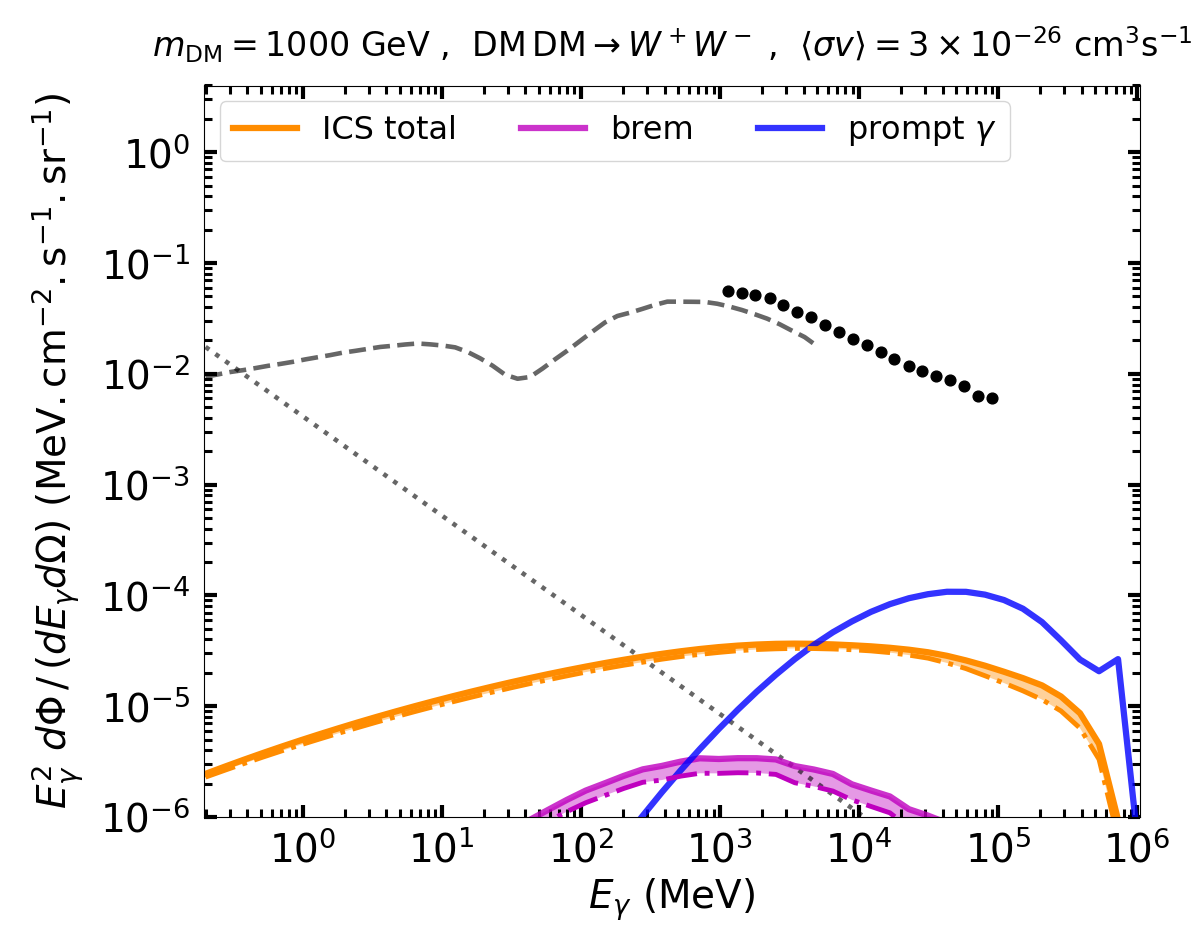}
\caption{\em {Same as Fig.~\ref{fig:ISRF_variation}, but 
considering other annihilation channels and other DM masses: 
$DM\,DM \, \rightarrow \, e^+\,e^-$, $m_{\rm DM} = 1$ GeV ({\it top left}); 
$DM\,DM \, \rightarrow \, e^+\,e^-$, $m_{\rm DM} = 100$ GeV ({\it top right});
$DM\,DM \, \rightarrow \, b\,\bar{b}$, $m_{\rm DM} = 10$ GeV ({\it middle left});
$DM\,DM \, \rightarrow \, b\,\bar{b}$, $m_{\rm DM} = 100$ GeV ({\it middle right});
$DM\,DM \, \rightarrow \, W^+\,W^-$, $m_{\rm DM} = 200$ GeV ({\it bottom left});
$DM\,DM \, \rightarrow \, W^+\,W^-$, $m_{\rm DM} = 1$ TeV ({\it bottom right}).}
}
\label{fig:ISRF_variation_2}
\end{figure*}

{In figs.~\ref{fig:ISRF_variation} and \ref{fig:ISRF_variation_2} 
we show the level of variation in the DM induced secondary fluxes 
due to two different choices for the ISRF model. 
We show such variations for different values of the DM mass and for 
different annihilation channels. 
Fig.~\ref{fig:ISRF_variation} shows these variations for the $\mu^+\mu^-$ channel 
considering three values of the DM mass, $m_{\rm DM} = $1, 10 and 100 GeV. 
Fig.~\ref{fig:ISRF_variation_2} shows the results for 
other channels and other $m_{\rm DM}$ values: 
$DM\,DM \, \rightarrow \, e^+\,e^-$, $m_{\rm DM} = 1$ GeV ({\it top left}); 
$DM\,DM \, \rightarrow \, e^+\,e^-$, $m_{\rm DM} = 100$ GeV ({\it top right});
$DM\,DM \, \rightarrow \, b\,\bar{b}$, $m_{\rm DM} = 10$ GeV ({\it middle left});
$DM\,DM \, \rightarrow \, b\,\bar{b}$, $m_{\rm DM} = 100$ GeV ({\it middle right});
$DM\,DM \, \rightarrow \, W^+\,W^-$, $m_{\rm DM} = 200$ GeV ({\it bottom left});
$DM\,DM \, \rightarrow \, W^+\,W^-$, $m_{\rm DM} = 1$ TeV ({\it bottom right}). 
These two figures serve as the representative of all the different 
DM annihilation scenarios considered in the present work.} 
In each of these plots, the solid curves correspond to 
the ISRF model discussed in 
sec.~\ref{sec:secondary_ICS} (which is used in the main results) and 
are the same as the ones shown in fig.~\ref{fig:DM_fluxes} 
{(for the $\mu^+\mu^-$ channel)}. 
On the other hand, the dashed-dotted curves 
correspond to the  ISRF model 
``Porter2006'' (see \cite{Evoli:2016xgn} for a detailed discussion) 
that uses the \texttt{GALPROP} 
dataset~\cite{2012ApJ...750....3A, 2012ApJ...752...68V} 
and provide an alternative modeling of 
the radiation fields~\cite{Porter:2005qx}. 
{Figs.~\ref{fig:ISRF_variation} and \ref{fig:ISRF_variation_2} 
show that, using this alternative ISRF model, 
the secondary fluxes and hence the total signals change by small amounts. 
Such a variation in the total signal in our energy range of interest 
is visible mostly for lower DM masses.  
With increasing values of the DM mass, this effect becomes less important. 
This statement is true for all representative 
annihilation channels considered in this work.}
Note from the discussion in sec.~\ref{sec:DMsignal} 
that the ISRF densities enter into the signal calculation 
through the ICS emission power as well as 
through the energy loss of $e^\pm$. 


\subsection{Magnetic field model}
\label{sec:appendix_MF_model}

\begin{figure*}[ht!]
\centering
\includegraphics[width=0.5\textwidth]{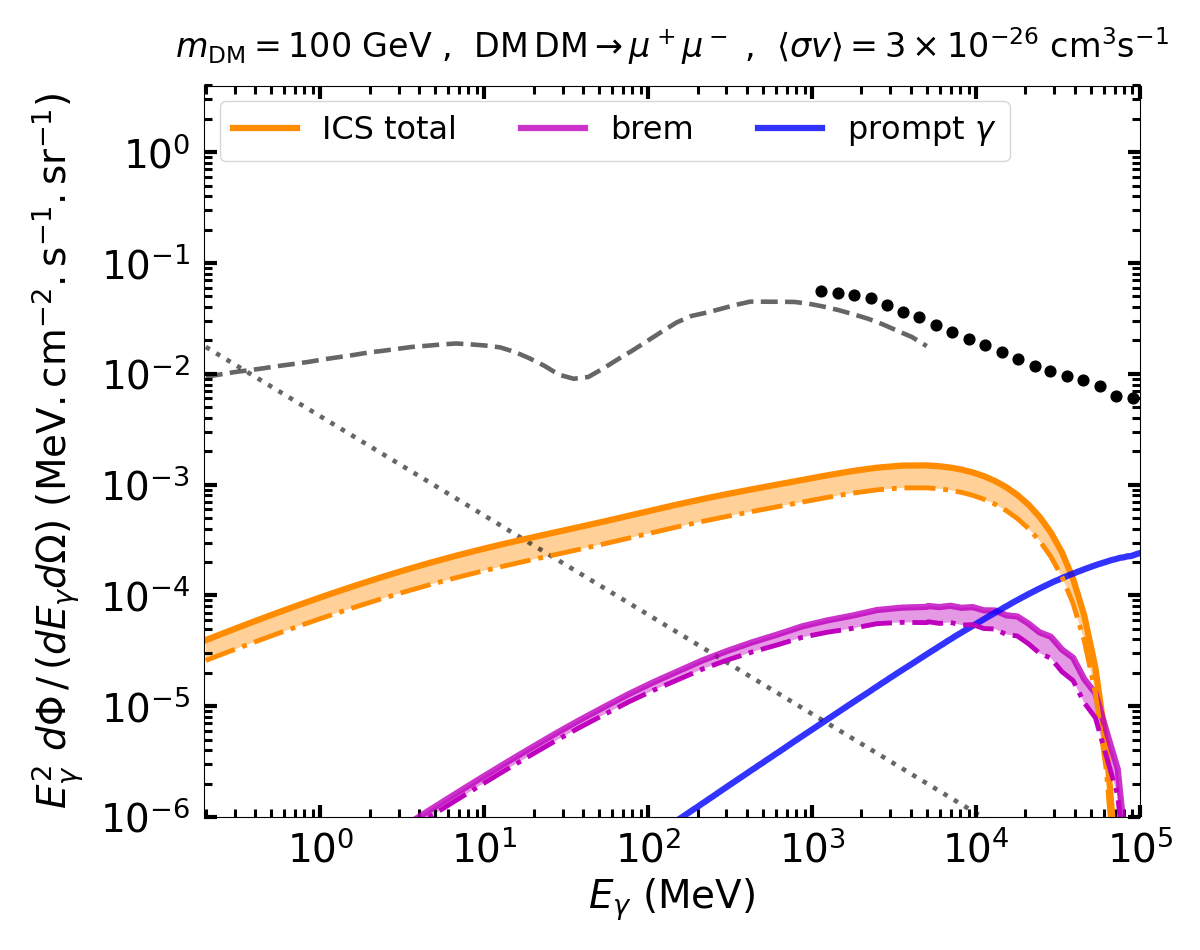}
\vspace{-2mm}
\caption{\em DM-induced signals considering different {\bfseries galactic magnetic field} models. 
The dashed-dotted curves (for the ICS and bremsstrahlung signals) 
correspond to the `MF3' model, that is comparatively 
stronger than the  `MF1' model used for the solid curves here and in the main results. 
See \cite{Buch:2015iya} for a detailed discussion on these 
magnetic field models.}
\label{fig:MF_variation}
\end{figure*}

Fig.~\ref{fig:MF_variation} shows the impact of considering two different Galactic magnetic field models, 
focusing again on the DM-induced secondary signals in the $\mu^+\mu^-$ annihilation channel for $m_{\rm DM} = 100$ GeV. 
The solid curves correspond to the model 
``MF1" (used in the main results) and are 
the same as  
the ones presented in fig.~\ref{fig:DM_fluxes}, 
while the dashed-dotted curves correspond to the model ``MF3" which 
has a relatively larger strength in the magnetic field. 
See \cite{Buch:2015iya} for a detailed discussion on 
these models. 
The latter model causes a (modest) weakening of the DM secondary signal up to a factor of 2, 
which can be understood as the fact that the $e^\pm$ lose more energy to synchrotron radiation 
in the stronger magnetic field (see the discussion in sec.~\ref{sec:DMsignal}). 

\subsection{{Galactic gas maps}}
\label{sec:appendix_gas_map} 

\begin{figure*}[ht!] 
\includegraphics[width=0.5\textwidth]{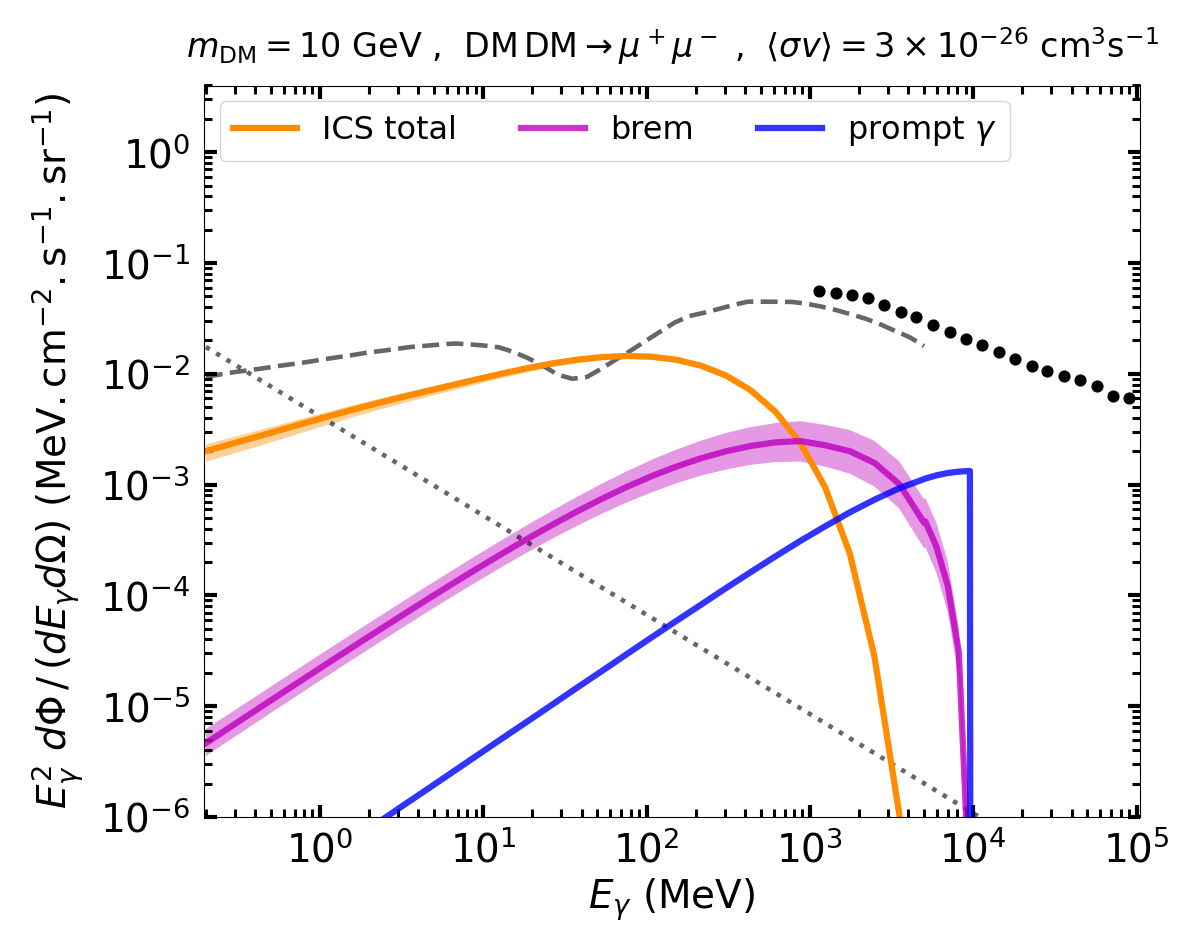}
\includegraphics[width=0.5\textwidth]{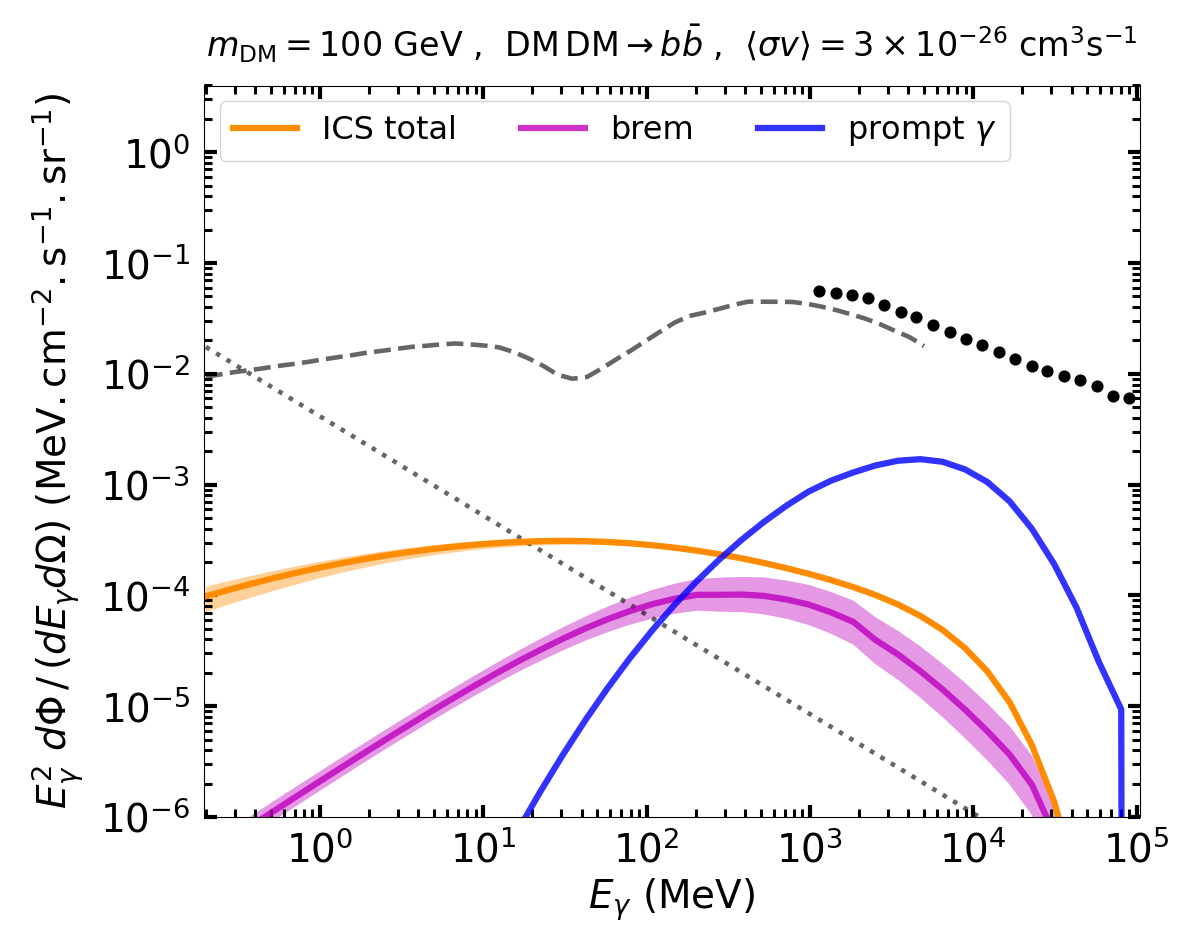}
\caption{{\em DM-induced photon signals considering a factor of 2 
{\bfseries variation} (in both directions) {\bfseries in the density maps of 
all Galactic gas particles}. 
The corresponding variations in different secondary photon signals are 
indicated by the associated bands. 
These are shown for $\mu^+\mu^-$ annihilation channel with $m_{\rm DM} = 10$ GeV (left panel), 
and $b\bar{b}$ annihilation channel with $m_{\rm DM} = 100$ GeV (right panel).
The solid curves in each case correspond to the signals 
obtained following Sec.~\ref{sec:DMsignal}.}}
\label{fig:gas_map_variation}
\end{figure*}

{In Fig.~\ref{fig:gas_map_variation} we show the level of changes in 
different secondary photon signals due to the variation of 
the Galactic gas densities. 
These are shown for $\mu^+\mu^-$ annihilation channel with 
$m_{\rm DM} = 10$ GeV (left panel), and $b\bar{b}$ annihilation channel 
with $m_{\rm DM} = 100$ GeV (right panel).
For illustration, we vary the densities 
of all gas particles by a factor of 2 above and below their values 
in the base model (mentioned in Sec.~\ref{sec:secondary_brem} 
and used in our main results). 
The corresponding variations in different photon signals are 
indicated by the associated bands. 
The solid curves in each case correspond to the signals 
obtained using the base model. 
As expected, the main noticeable changes occur 
in the bremsstrahlung signal since its production 
relies on the density of the gas targets. There is also a 
very mild variation in the ICS signal through the 
energy loss term of $e^\pm$ that depends 
(through bremsstrahlung, Coulomb interaction and ionization processes) 
on the target gas densities. As can be seen from Fig.~\ref{fig:gas_map_variation}, 
the DM induced total photon signals for different cases remain 
almost robust against the above-mentioned changes in the gas density profiles.}  

\newpage 

\subsection{Atmospheric backgrounds}
\label{sec:atmospheric_backgrounds}

\begin{figure*}[ht!]
\centering
\includegraphics[width=0.5\textwidth]{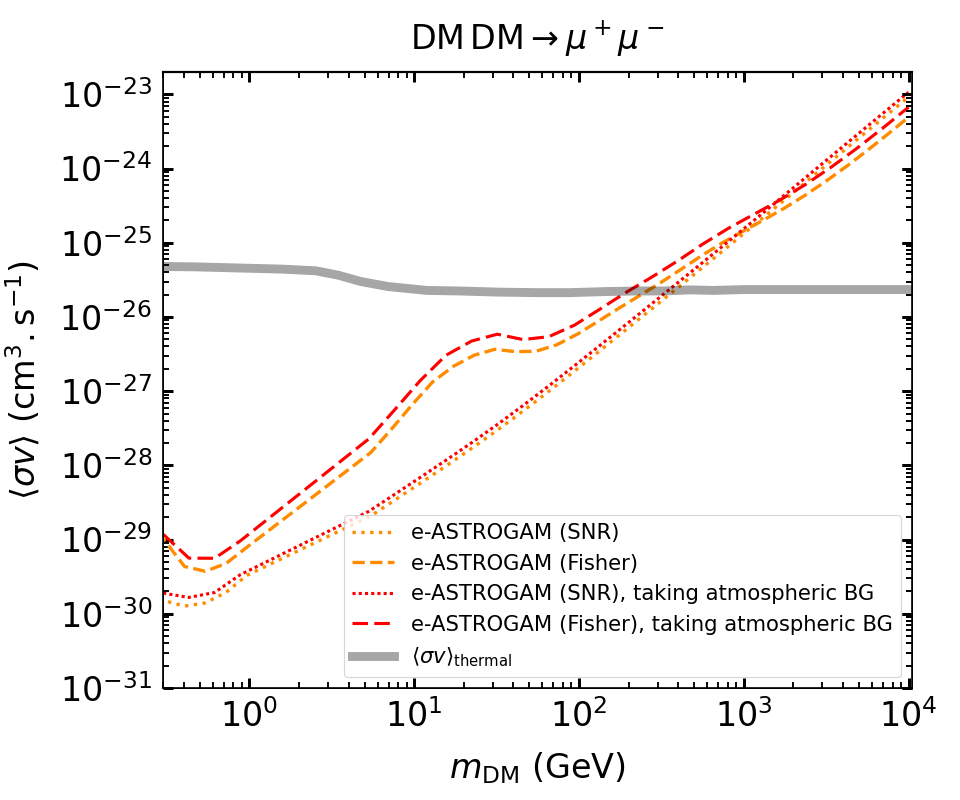}
\vspace{-1mm}
\caption{\em Effects of the inclusion of the possible 
{\bfseries atmospheric background} for {\sc e-Astrogam}. 
The red curves are obtained by taking into account 
the atmospheric background estimated in \cite{e-ASTROGAM:2016bph}, 
while the orange curves do not consider this and 
are the same as the ones  
shown in fig.~\ref{fig:sv_mx_limits_1}. 
}
\label{fig:sv_mx_atmosphericBG}
\end{figure*}

Among different future telescopes considered in this work, 
only {\sc e-Astrogam} so far provides an estimate for the 
possible atmospheric background 
(see figure 18 of \cite{e-ASTROGAM:2016bph}).  
Such a background may arise due to various phenomena in the Earth's atmosphere 
including the atmospheric lightnings and thunderstorms.
In fig.~\ref{fig:sv_mx_atmosphericBG} we show an example of the 
effects of the inclusion of this atmospheric background in the analysis. 
For the methodology to include such a background we follow \cite{ODonnell:2024aaw}. 
As can be seen, for {\sc e-Astrogam} the effects of the inclusion of the 
atmospheric background should be mild. 

\newpage 

\bibliography{bibliography.bib}

\end{document}